\documentclass{article}
\usepackage{graphicx}
\usepackage{setspace}
\usepackage{subcaption}
\usepackage{placeins}
\usepackage{authblk}
\usepackage{array,float,amsmath}
\newcolumntype{C}[1]{>{\centering\let\newline\\\arraybackslash\hspace{0pt}}m{#1}}
\usepackage{hyperref}
\usepackage[margin=1in]{geometry} 

\newcommand{\conf}[3]{$#1$ (95\% CI: $#2$--$#3$)}

\title{Simulating COVID-19 in a University Environment}

\author[1]{Philip T. Gressman}
\author[2]{Jennifer R. Peck}
\affil[1]{Department of Mathematics, University of Pennsylvania}
\affil[2]{Department of Economics, Swarthmore College}

\begin{document}
\maketitle

\begin{abstract}
  \noindent Residential colleges and universities face unique challenges in providing in-person instruction during the COVID-19 pandemic.  Administrators are currently faced with decisions about whether to open during the pandemic and what modifications of their normal operations might be necessary to protect students, faculty and staff.  There is little information, however, on what measures are likely to be most effective and whether existing interventions could contain the spread of an outbreak on campus.  We develop a full-scale stochastic agent-based model to determine whether in-person instruction could safely continue during the pandemic and evaluate the necessity of various interventions. Simulation results indicate that large scale randomized testing, contact-tracing, and quarantining are important components of a successful strategy for containing campus outbreaks.  High test specificity is critical for keeping the size of the quarantine population manageable.  Moving the largest classes online is also crucial for controlling both the size of outbreaks and the number of students in quarantine.  Increased residential exposure can significantly impact the size of an outbreak, but it is likely more important to control non-residential social exposure among students.  Finally, necessarily high quarantine rates even in controlled outbreaks imply significant absenteeism, indicating a need to plan for remote instruction of quarantined students.
  
  \end{abstract}

  \onehalfspacing

  \section{Introduction}

In March of 2020, most American colleges and universities closed to in-person instruction, moving classes online and sending nearly all residential students home.  In the coming fall, as many as 17 million of these graduate and undergraduate students could return to campuses, over two million of whom would likely live on campus \cite{des}.\footnote{In 2015-2016, 10.8 percent of undergraduates and 27.3 percent of graduate students took all of the courses for their degrees online \cite{des}.  Approximately 16\% of U.S. undergraduate postsecondary students live in university-owned housing.}
  In the midst of the global COVID-19 pandemic, this presents an unprecedented public health and logistical challenge.  At the same time, reopening to in-person instruction is viewed as extremely important both pedagogically and financially \cite{paxson}.   The experience of many institutions in the spring reveals that \textit{on-campus} instruction is important even if it doesn't all occur in person: there were widespread reports of students experiencing limited access to computing resources, intermittent internet access, and even homelessness \cite{homeless}.  In providing these resources, residential colleges facilitate both instruction and appropriate evaluation of student learning.

Residential universities face a unique challenge in balancing the benefits of on-campus and in-person instruction with the risks and costs of operating during the pandemic.  One the one hand, universities are extremely constrained in their ability to control the spread of a virus on campus: students, faculty and staff all have repeated, regular personal interactions in a closed space, and institutions have limited resources to test and quarantine students.  At the same time, administrators also have an extremely high degree of control over the parameters of many of these interactions, and can set what students are in which classes, when those classes meet, and who students interact with in a residential setting.  Because of these unique features, standard epidemiological models are of limited value for assessing the spread of a COVID-19 outbreak on a college campus and for identifying the tools administrators could most effectively use to minimize the likelihood and magnitude of such an outbreak.
Many residential colleges and universities around the country and the world are therefore currently facing critical decisions about how to adapt their normal operations to protect students, faculty, and staff from COVID-19 during the coming academic year. An April 22nd \textit{Inside Higher Ed} article \cite{ihe} lists 15 different possible approaches to the problem, each with clear strengths and weaknesses. In the absence of relevant prior experience, these institutions are largely in the dark about how one might expect a COVID-19 outbreak to evolve in the unique environment of a college campus and how much of an effect the many possible mitigation strategies should be expected to produce.

To address the absence of reliable evidence, we have we have endeavored to create a simulation which is as realistic as is reasonably possible. Our simulation\footnote{Source code is available online at \url{https://www.github.com/gressman/covid_university}.} is a full-scale stochastic agent-based model (ABM) of a reasonably large research university. In it, we are able to study the spread of a highly virulent illness like COVID-19 and monitor the efficacy of tools like quarantining and contact tracing, mask-wearing policies, partial transitions to online instruction, and others.  Agent-based models such as ours are uniquely well-suited to the situation we study because behavior of both instructors and students in an academic environment is highly structured spatially and temporally, which can lead to consistent overestimates of disease incidence when other tools like structured metapopulation models are used \cite{ajelli}. The modest scale (22,500 total agents) of a university community also means that concerns about computational complexity of ABMs like ours are not applicable.

It is important to emphasize that despite the many realistic features we have incorporated, any such simulation is, by its nature, a fundamentally limited caricature of a complex real-world situation. However, there are some key lessons that we believe can be reasonably carried over to inform real-world decisions:

\noindent
\begin{itemize}
\item \textbf{Testing accuracy is a critical issue which needs immediate attention.} In an aggressive contact-tracing scenario, it is reasonable to expect every positive test result to lead to the quarantining of 10--20 students.  In a population with a low infection rate, a high quarantine rate can still result if testing is not extremely high accuracy.  In particular, the most important driver of quarantine population is the false positive rate of the test. The functional difference between tests with 1\% and 0.01\% false positive rates are enormous in the university setting, but none of the existing molecular tests have been examined closely enough to carefully determine the order of magnitude of the false positive rate. At least a few have been shown to have rates higher than 1\%, which could lead to the unnecessary quarantining of literally thousands of students over the course of a semester.

\noindent
\item \textbf{Holding large classes greatly increases the risk of a significant outbreak on campus.} Our simulation finds that all reasonably successful strategies involve moving instruction for large classes entirely online. In particular, dividing large classes into smaller sections which still physically meet does not appear to have a strong enough effect to protect students and instructors.

\noindent
\item \textbf{It is extremely important that students refrain from all contact outside of academic and residential settings.} Even very small rates of contact in large group settings like dining halls or parties may be sufficient to sustain an outbreak on campus regardless of any other protective measures which have been put into place.

\noindent
\item \textbf{All instructors need to prepare for extended student absences due to quarantine.} In our best-case scenario, classes spend an average of 1-2 weeks of the semester with at least 10\% of the students absent due to quarantine. For this reason, we expect that even small classes which are permitted to meet in person will still need procedures in place for continuity of instruction for quarantined students.
\end{itemize}

Our study is a complement to recent work by Weeden and Cornwell on how large enrollment networks might contribute to the spread of disease on campus \cite{weeden}.  They show that students are highly interconnected through their courses and argue that the ``small world'' nature of enrollment networks makes the student population susceptible to high rates of disease transmission.  Removing large courses from the university network (by moving them online) decreases the connectivity in these networks, but is not by itself enough to contain outbreaks.  Our work examines these enrollment networks with simulated rather than empirical network data, and includes instructor interactions, models residential exposure, and allows for non-residential exposure.  We also include heterogeneity in the amount of contact that students experience with one another in classes and account for temporal variation as courses meet throughout the week.  Incorporating current best-known immunological and epidemiological parameters also allows us to examine how different intervention measures might complement one another in slowing disease transmission. Because the specific question about the effect of moving large classes online is so important, we also include a simpler analytical model in the Appendix which reinforces what we see in our ABM.

The remainder of the paper describes the structure of our ABM and discusses some of its most significant parameters (Section \ref{methods}).  We then present some key implications for the spread of infection and the size of the quarantine population corresponding with different control methods (Section \ref{results}).  Section \ref{conclusion} concludes.

\section{Methods}
\label{methods}
In this section we give a description of the overall structure of our model and highlight several of the more significant parameters. Our choice of ABM to approach this problem is motivated by several factors. First and foremost, ABM allows us to incorporate very precise information about contact heterogeneity between populations (instructors and students) and even among students at different stages of their university careers. It is well-known that contact heterogeneity can significantly affect dynamics \cite{sattenspieldietz,meakinkeeling} as can temporal contact structure \cite{seno}, which are both fundamental features of interaction in a university setting. In situations when behavior is highly structured, it is known that maintaining the identity of individuals is important \cite{kdvh,ajelli}. For these reasons, ABMs are uniquely well-suited to small-scale problems such as ours and have already been used in
studies of COVID-19 \cite{az,arXiv1}.

\subsection{Viral Dynamics}
\label{dynamics}
Each individual has a state which reflects their health status: Susceptible (S), Infected (I), or Removed (R). Infected individuals have refined states which reflect the incubation period they will experience, whether or not they are symptomatic, and how many days their infection has progressed. Independently of health status, individuals may also be quarantined; when quarantined it is assumed that susceptible individuals cannot become infected and infected individuals cannot transmit infection to others.

The sequence of events in a simulated day is straightforward:
\begin{enumerate}
\item Illness Testing: A predetermined fraction of the population is randomly selected for testing. The default value is set at 3\%, which means that members of the community will be tested approximately once per month. Anyone who was flagged via contact tracing on the previous day is also tested.
\item Quarantining: Anyone from the Illness Testing step whose test results were positive \textit{and} any symptomatic infected individuals who develop symptoms on this day are quarantined immediately if not already quarantined. Additionally, such individuals are flagged to have their recent contacts traced.
\item Status Updates: Individuals who have been quarantined for 14 days are released. If they were susceptible when entering quarantine, they emerge as still susceptible; those who were infected or removed upon entering quarantine emerge as removed. Similarly, individuals who have been infected for 14 days are updated to removed (R) status. We note that while there is some evidence to suggest that a small fraction ($\sim$$1\%$) of infected individuals become symptomatic 14 days or more after infection \cite{incubation1}, we neglect this possibility. 
\item Contact Tracing: Every individual who was flagged earlier in the day (due to testing positive or developing symptoms) has their contacts traced.  This includes all of their (non-quarantined) contacts from the previous two days (not including today, and not including days when the flagged individual was quarantined).\footnote{The choice of two days is consistent with current CDC guidelines (revised May 29, 2020) on the contact elicitation window.  This protocol matches the recommendations of the WHO, European CDC, and Public Health Canada. 
}   These contacts are flagged for testing tomorrow and are quarantined immediately.  The purpose of testing these individuals after they have already been quarantined is to trace \textit{their} contacts (after the test on the next day) if the test is positive.
\item Infection Transmission: Non-quarantined susceptible individuals who contact non-quarantined infected individuals become infected themselves with a probability that depends on the infection state of the infected. Additional details about the model of transmission dynamics appear below.
\item Outside Transmission: On any particular day, there is a 25\% chance that one non-quarantined susceptible individual becomes spontaneously infected due to presumed transmission from non-university contact. By the state-level standards in the ``Process to Reopen Pennsylvania,'' this rate is rather low, roughly $1/3$ of the amount which allows transitioning from the most restrictive ``Red'' phase to the ``Yellow'' phase (which corresponds to at most 50 new cases total per 100,000 in a 14-day period). Thus our default scenario models a relatively low level of transmission outside the university. Whether this low level can be achieved broadly in the U.S. by late August is a question of growing concern; results based on higher levels of outside transmission are summarized in Tables \ref{outside1} and \ref{outside2} in the Appendix.
\end{enumerate}

The incubation period for each infected individual is fixed to have mean 5.2 days and is randomly assigned to an individual upon infection. The exact distribution used is a discretized Gamma distribution with shape parameter $k=4$ (details are in Section \ref{gamma} of the Appendix).  The overall mean of 5.2 days is consistent with several estimates in the literature \cite{incubation1,early}.

The infectiousness of an infected individual is a function of the time since infection. The serial interval is based on an estimate of 5.8 days  \cite{serialinterval} by setting transmission probability, which is the probability that any one \textit{potential} transmission contact actually results in infection, to be proportional to probability density function of a discretized Gamma distribution with mean $5.8$ and shape $k=4$.  Transmission probabilities are unknown parameters, so we calibrate the model in such a way that the basic reproduction number $R_0$ for non-residential contacts roughly matches the most relevant estimates in the literature.  In particular, we normalize transmission probabilities based on expected rates of contact (described in the next section) so that an infected individual not subject to quarantine would transmit the disease to an average of 3.8 individuals over the course of the illness. In other words, probabilities have been adjusted so that when residential effects are ignored, the basic reproduction number $R_0$ for the simulation is effectively $3.8$.  As with the expected number of contacts, it is important to emphasize that \textit{the actual reproduction number will depend on the circumstances being simulated}: as various prevention measures are taken, the value of $R_0$ decreases relative to a scenario where all contacts are susceptible and no measures are taken to combat illness.  While the earliest analyses of the Wuhan outbreak estimated $R_0$ to be closer to $2.2$, subsequent work tends to point towards somewhat higher values. For example, Sanche \textit{et al.} estimate $R_0$ to be \conf{5.7}{3.8}{8.9} \cite{r0high} and Flaxman \textit{et. al.} find a similar value of \conf{3.87}{3.01}{4.66} \cite{imperial1}. Given the reasonable expectation that contact rates on college campuses will be higher than the general population, it is natural to assume that these higher estimates are more appropriate for the current modeling purposes.  This is consistent with estimates of $R_0$ for influenza outbreaks \cite{metaflu}, which tend to be higher in confined settings like schools, military bases, and ships.\footnote{A review of the literature on $R_0$ estimates for influenza \cite{metaflu}  finds generally higher point estimates for $R_0$ in confined settings in all but the 1968 pandemic.  Median $R_0$ point estimates in confined settings were 3.82 for the 1918 pandemic and 1.96 for the 2009 pandemic compared with 1.80 and 1.46 in community settings.}
We have also investigated the stability of our results with respect to alternate choices of $R_0$ and have summarized the results in Tables \ref{outside1} and \ref{outside2} in the Appendix.

Additional features of the dynamical aspects of the simulation are as follows:

\begin{itemize}
\item We assume that 75\% of those infected have cases which are mild, asymptomatic, or simply unidentified. These individuals do not spontaneously self-report for quarantine in Step 2 above; they are modeled as half as infectious as symptomatic counterparts \cite{asymptomatic}.
\item There is a global parameter to account for a variety of nonpharmacological interventions (e.g., mask-wearing). The default value is chosen to so that transmission probabilities are reduced to $50\%$ of the values identified above. This decreased transmission is included reflect the expected effects of universal mask-wearing policies. MacIntyre \textit{et al.} \cite{masks} find that masks in the home reduce transmission rates to roughly $30\%$ of their otherwise-expected values, but caution against extrapolating their results to repeat-contact settings such as this one. Because our model explicitly tracks differences between individual and repeat contacts, it seems reasonable to expect that every individual transmission event in our setting will see roughly similar benefits to those found in the home.
\item We assume that on the first day of classes, $5\%$ of the community has already obtained immunity (and hence reside in the R state). These individuals are not assumed to be known to the administration, and consequently are subject to quarantine if they have been identified to have recent contact with an individual known to be infected.
\end{itemize}

Each simulation is initialized with zero infected individuals and runs for 100 days.

\subsection{False Positive Rate (FPR)}

One feature which is critical to the dynamics of quarantine populations is false positive results during the Illness Testing phase.
We assume that the false positive rate (FPR) is $0.1\%$ and the false negative rate is $3.0\%$ in the main analysis. We have endeavored to select these parameters to be consistent with what is possible with existing tests \cite{testing}. Tables \ref{fpr1} and \ref{fpr2} also show results for a wide range of values of the FPR and FNR; there is additional discussion of the FPR in Section \ref{results}. One can see from the tables that outcomes do not depend sensitively on the FNR but can vary by orders of magnitude under different values of the FPR.

It is extremely difficult to identify a reasonable value for the FPR given the current data on available tests.  The clinical evaluations reported in Emergency Use Authorizations (EUAs) for many molecular tests currently approved by the FDA involve testing only 30 negative reference samples.  This sample size is too small to guarantee an FPR of much below $3\%$, which in our simulation leads to catastrophically high rates of quarantine. At the same time, however, very high specificity (i.e., a lower FPR) is not \textit{ruled out} by the absence of testing. To arrive at a reasonable order of magnitude, we consulted results from the Foundation for Innovative New Diagnostics, which established an average clinical specificity of 99.4\% for a number of the most widely-used molecular tests \cite{FIND}; these results are considered tentative, as it is perhaps possible that they are a function of improperly defined negative reference standards (i.e., the level at which a result is considered negative) rather than true false positives.  We also reviewed 61 EUAs approved by the FDA for molecular tests and found six that reported false positive test results with small sample sizes ($\sim$30) \cite{FDA}. Thus it may be reasonable to expect that a realistic FPR may be on the order of $0.3\%$--$0.6\%$. Scholarly work on the FPR of COVID-19 testing is extremely rare, but once source makes a conservative estimate of the FPR at $0.8\%$ based on external quality assessments of similar assays \cite{fpr}.

A simple calculation illustrates a serious problem with this FPR: in a community of 22,500 people, $3\%$ daily testing means conducting $675$ tests per day. A false positive rate of $0.8\%$ means that an average of $5.4$ tests per day will result in false positives assuming that the prevalence of true infections is small or zero. In isolation this is a modest number; however, when combined with contact tracing, $5.4$ false positives per day is clearly infeasibly high when one considers that $10$--$20$ individuals will be quarantined for each (falsely) positive individual. This means that a university that is entirely disease-free would send $54$--$108$ people \textit{per day} into quarantine. If individuals are kept in quarantine for $14$ days, this means a peak quarantine population would be 750--1,500 in quarantine at any given moment (about $3\%$--$7\%$ of the university population). Over the course of the semester, 5,000--10,000 individuals would be sent into quarantine at one point or another, which is $22\%$--$44\%$ of the population, \textit{even in the absence of any true infections}.

In short, a naive approach to mass testing in a university environment can be reasonably expected to quickly lead to a crisis of over-quarantining. The good news is that such a scenario is entirely avoidable with proper planning. By developing testing protocols under which (for example) positive samples are always retested, the FPR can be effectively lowered to a more sensible level. Pooling samples may also allow for reasonable retesting strategies at minimal cost.  As the benefit of such protocols so vastly outweighs this cost, we assume that reasonable institutions will do so and set the FPR for the testing protocol in our main simulation to $0.1\%$.

The selection of a FNR of 3\% is also consistent with values reported in many EUAs. When the overall prevalence is illness is low, we find that the disease and quarantine dynamics are not significantly affected by substantial variation of this parameter (see the Appendix): when the total number of infected individuals is below $100$ for the entire semester (as it is in a number of scenarios we model), the number of false negative results encountered is correspondingly small--only 2 or 3 total in a semester. This is vastly different than the case of the FPR when thousands of true negative tests can lead to hundreds of false positive results.

\subsection{Contact Patterns}
\label{contactpatterns}
 Every run of the simulation consists of $20,000$ students and $2,500$ instructors who interact daily for 100 days. Every day, simulated individuals experience contacts with each other, and these contacts are the basis for both disease transmission and the mitigation strategy of contact tracing. We allow for contacts to be asymmetric (so that the likelihood of transmission from person $a$ to person $b$ need not necessarily be the same as the likelihood of transmission in the reverse direction) but in almost all situations we model both directions of transmission as equally likely. We also categorize contacts as traceable or nontraceable. This latter category is meant to model contacts which are incidental or otherwise not sufficiently memorable to be recalled upon reflection.
  
 There are two main sorts of contacts in the simulation. The most common sort of contact is \textit{Poisson contact}, generated by a Poisson point process whose rate is determined by a detailed scoring system which takes into account whether the two individuals have any common activities on a given day and the nature of their roles in those activities (e.g., a student and an instructor in a class will have a higher rate of contact than two random students in the same class, but lower than the rate of contact of two friends in the class). Activities include classes, recitations, department memberships, and friendships (which are formed at the beginning of the simulation between smaller groups of individuals within larger classes, recitations, and departments).  These contacts model a wide variety of possible scenarios that one might encounter in a university setting. Details can be found in Section \ref{appendix:contacts} of the Appendix. In particular, there are several key ``scales'' on which contacts occur: a close contact scale which captures interactions like friendships and nearest classmates in a classroom, a classroom scale which includes all individuals in a room at the time of class, environmental contacts that occur around physical presence in a building or department (like shared elevators or hallways), and broad environmental contacts which can occasionally occur between any two individuals on campus. We model contacts on each of these four scales as roughly equally likely to occur, and consider the close contact and classroom scale contacts to be traceable and the department and broad environmental contact scales to be nontraceable. For convenience, we refer to all the forms of contact above as ``academic contacts'' to emphasize that they are all heavily influenced by the overall rate of in-person (as opposed to online) academic activity. We also include a category of broad social contacts within the Poisson contact system to capture contacts at a large social scale which are not directly tied to the acts of going to or from classes.
 
The second type of contact is \textit{residential contact}, which is guaranteed to occur every day. Residential contacts are \textit{de facto} stronger than other contacts in the sense that transmission in both directions is always allowed (whereas under normal circumstances there is effectively only a $50\%$ chance of a particular direction of transmission being allowed for a Poisson contact.) As a robustness check, we constructed two distinct models of residential contacts: a ``discrete'' model in which students are assembled into permanent, small residential groups of fixed size, and a ``linear'' model in which students are effectively arranged in linear dormitories and have permanent contact with small numbers of adjacent students. Both models gave very similar results, so we have adopted the linear model as standard. We assume under normal circumstances that students have an average of $1$ residential contact, which means that the typical residential arrangement is that students are housed in pairs (i.e., an average student has one daily residential contact with some other student).  We also test the effects of increasing the number of these types of residential (and other) contacts.
 
 It is important to emphasize that the average daily number of contacts for a simulated individual \textit{is a function of many factors and is not an adjustable parameter}. Our contact system is normalized to yield an average of approximately $11$ traceable and $8$ nontraceable contacts per person per day when all classes are meeting physically and no social distancing is being exercised. This is consistent with a number of surveys of the general public which found that participants' self-reported average daily contacts fell in the range of 10 to 14 per day \cite{contacts,contacts2}.  Determining a realistic rate for nontraceable contacts is a bit more challenging, but we note that under our normalization, the attack rate in a residential setting (i.e., the rate of transmission to individuals contacted every day) is $32.9\%$, which is similar to the observed secondary attack rate to spouses in a household setting \cite{household}.  We examine the sensitivity of the results to the total number of daily contacts in Tables \ref{appendix:contact1} and \ref{appendix:contact2} in the Appendix.
 We emphasize that the average student in our simulation has at least one class in common with 244 others on average, so the contact process necessarily assumes that it is rather unlikely that a student will experience contact with any particular one of those individuals on any given day. 
 
The number of courses is variable but is $3,750$ in expectation. Very large classes with more than 150 students are split roughly evenly into sections. Classes are independently randomly scheduled as Monday, Wednesday, Friday (40\%), Tuesday-Thursday (40\%), or Monday-Wednesday (20\%). 
 Classes of more than 50 students are also assigned teaching assistants who run additional once-per-week recitations with 20 students at once. Each assistant is also a student; any student who is not taking the given course or any other course of lower difficulty is eligible to be an assistant, and assistants are limited to be responsible for at most 80 students. Recitations meet only one day per week, and never on the day that the corresponding lecture is held.

Students are evenly distributed into 8 large cohorts. An individual's cohort affects the probability that they will enroll in a particular course, with students in cohort 0 (corresponding to freshmen) being most likely to select courses of lowest difficulty and students in cohort 7 (roughly, advanced graduate students) being most likely to select courses of highest difficulty. Each student enrolls in either 4 or 5 courses (with both being equally likely) The precise distributions for course preferences were designed in an ad hoc way to generate a distribution of class sizes which is consistent with reported data at the University of Pennsylvania. The distribution of class sizes for a typical randomly-generated university are summarized in Figure \ref{sizes}. Additional implementation details are given in Section \ref{appendix:enroll} of the Appendix. We also note a few important statistical features of our simulated contact process which are similar to the values found in the actual academic enrollment network at Cornell as reported in \cite{weeden}:
\begin{itemize}
\item The mean class size in our simulated university is $24$.
\item Roughly 90\% of classes have 50 or fewer students.
\item The graph-theoretic properties of our simulated university also closely mirror real-world data: students are, on average, connected $1.2\%$ of the entire student body through common courses (i.e., the average student has a common classmate with $244$ others).  The collection of classmates-of-classmates for an average student includes nearly half the student body (i.e., the average student is connected to $50.0\%$ of all others via a 1- or 2-edge path), and by the next step, $99.2\%$ of the student body can be reached via a path of length at most $3$. There are, on average, $2.5$ ``degrees of separation'' between a typical pair of students (i.e., the average geodesic distance in the classmates graph is approximately $2.5$).
\end{itemize}

\begin{figure}[h]
\begin{center}
\includegraphics[width=.7\textwidth]{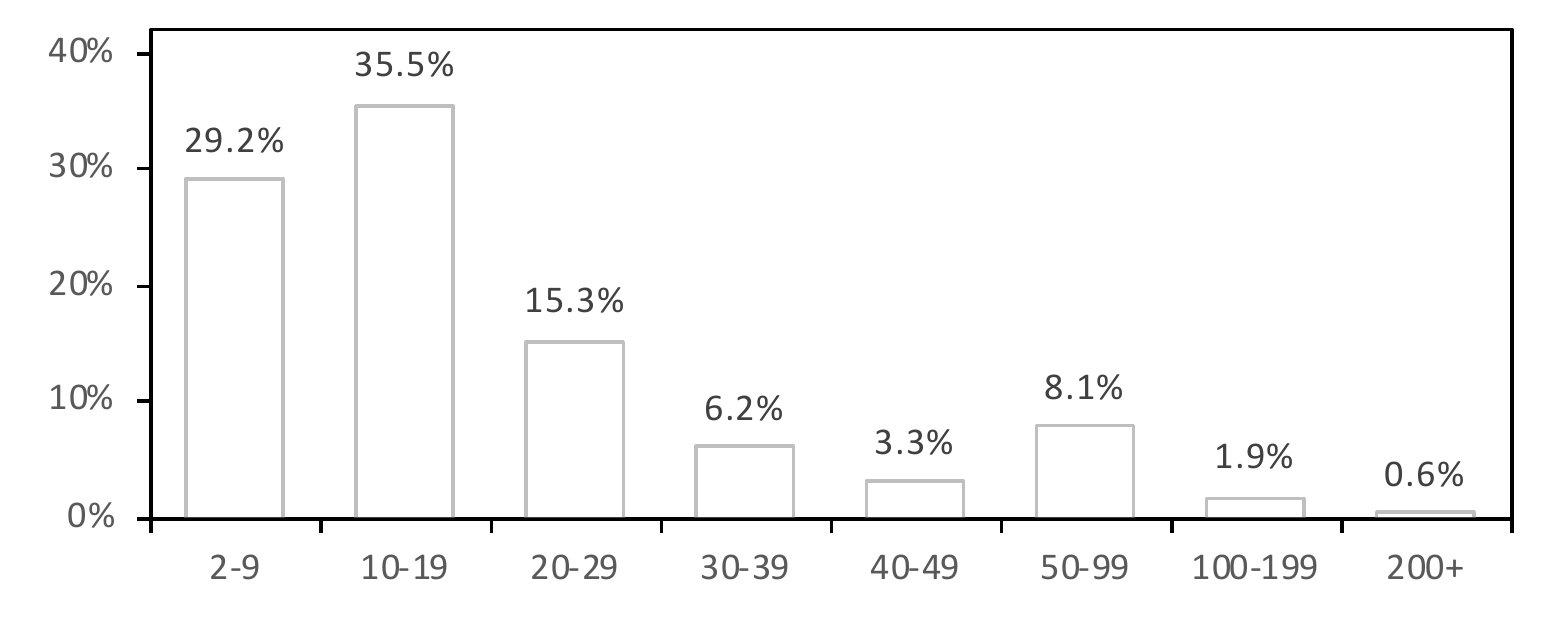}
\end{center}
\caption{Distribution of class sizes (before splitting very large classes into smaller sections) averaged over 100 randomly-generated universities. Here the proportion indicates the fraction of all $\sim$3,750 courses which fall into the given size bin. The largest class in the university has approximately 800 students before subdivision.}
\label{sizes}
\end{figure}

Within each class with at least 5 students, it is assumed that students make smaller friend groups (the so-called ``close contact'' scale from the previous section) with size that scales like the square root of the overall class size (e.g., a class of 100 students would be subdivided into roughly 10 social groups of 10 students each). This is meant to reflect natural social tendencies of students and to reflect the fact that students tend to arrange themselves in the classroom in similar ways from day to day, so that contacts inside a classroom environment are unlikely to be uniformly random. Friends are most likely to meet on the days when their common activity  is in session, but still have a smaller but positive rate of contact on other days as well.

Classes are also randomly assigned to one of 120 departments with frequency which is exponentially decaying, so that the largest department is expected to be ten times the size of the smallest department in terms of the number of courses offered. Once classes are assigned to departments, instructors are assigned to departments in proportion to the number of classes taught by each department (very small departments are guaranteed at least one instructor). Instructors within the same department have additional positive contributions to their Poisson contact rates during the week but not during the weekend. As with students, instructors also form friendships within their department.

\section{Results}
\label{results}

As a small validation of the model, we analyzed the early-phase doubling period for uncontrolled growth when no intervention scenarios are applied. Specifically, we compute the doubling period for cumulative infections over the range of days beginning with the first infection and ending on the day when at least 2000 individuals have been infected (note: all 500 simulated runs exceeded 2000 infected individuals). We found that the median early-phase doubling period for our model was $2.185$ days, with $50\%$ of simulations falling between $2.004$ and $2.365$ days. The observed doubling period of cases in the United States between March 1 and March 31 was 2.53 days \cite{CDC}. It is reasonable to expect the doubling period within an American university to be somewhat shorter than was observed in the general population. To develop a sense of what realistically short doubling periods might be, we observe that data from Quebec's Federal Training Centre prison showed a doubling period of approximately $1.94$ days between April 10 and April 21 \cite{prison}.

Our two main outcome variables are the total number of people infected and the peak number of students in quarantine during the semester.  These two outcomes summarize the risk to the campus community from the disease and the costs and academic disruption involved in having students in quarantine on campus.  Our baseline level of intervention (the ``standard intervention''\footnote{We have constructed the standard intervention as a tool to study the most widely-discussed and broadly-applicable on-campus infection mitigation strategies, but we note that there are many other potential strategies which institutions might develop around their own needs and capabilities. One such interesting idea is the formation of smaller, separated academic cohorts \cite{plavchan}, which may be able to capitalize on the structure of social networks to limit disease spread \cite{socialnetworks}.}) consists of the combination of quarantine and contact tracing, universal mask-wearing, daily randomized testing of 3\% of the university community, and transitioning all classes with 30 or more students to online-only interaction\footnote{We investigated several other possible interventions, including dividing large classes into many smaller sections and/or bringing only certain cohorts of students back to campus. These alternatives were found to be far less effective than moving large classes online, so we will not focus on them here.}. As part of the online transition, we model the social distancing which is afforded by moving smaller classes into the physical classrooms vacated by larger classes which have moved online.  It's worth noting that there are two main classes of intervention: those that reduce both infections and quarantine (mask-wearing, social distancing, transitioning large classes) and those that reduce infections but increase quarantine (quarantine and contact-tracing, randomized testing).  Increasing testing accuracy (reducing the FPR) would be expected to reduce quarantine without significantly affecting infections.

\subsection{Overview of Control Measures}

\begin{figure}[!ht]
\begin{center}
\begin{subfigure}[ht]{.5\textwidth}
\centering
\includegraphics[width=\textwidth]{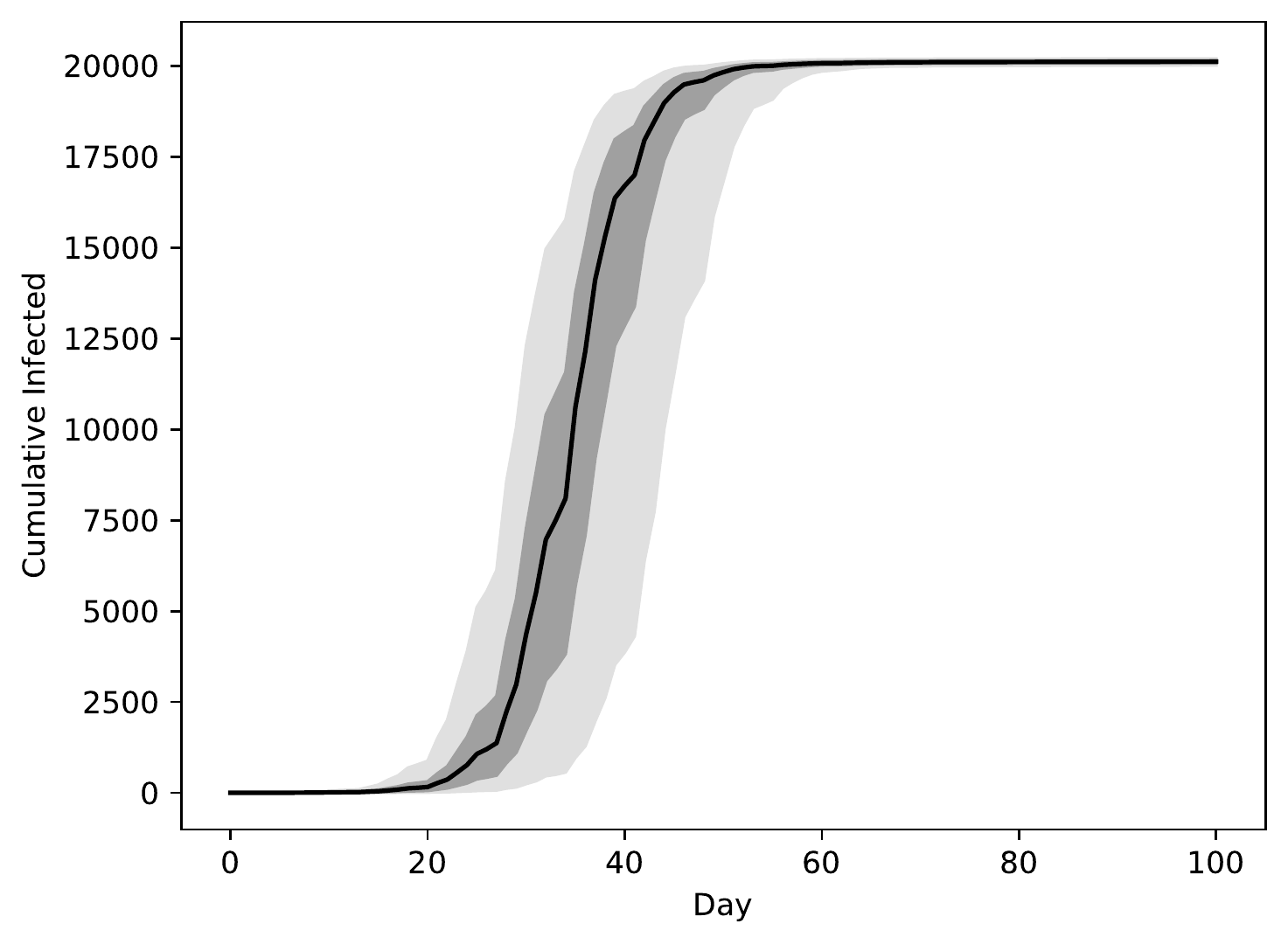}
  \caption{No Intervention}
\end{subfigure}%
  \begin{subfigure}[ht]{.5\textwidth}
\centering
\includegraphics[width=\textwidth]{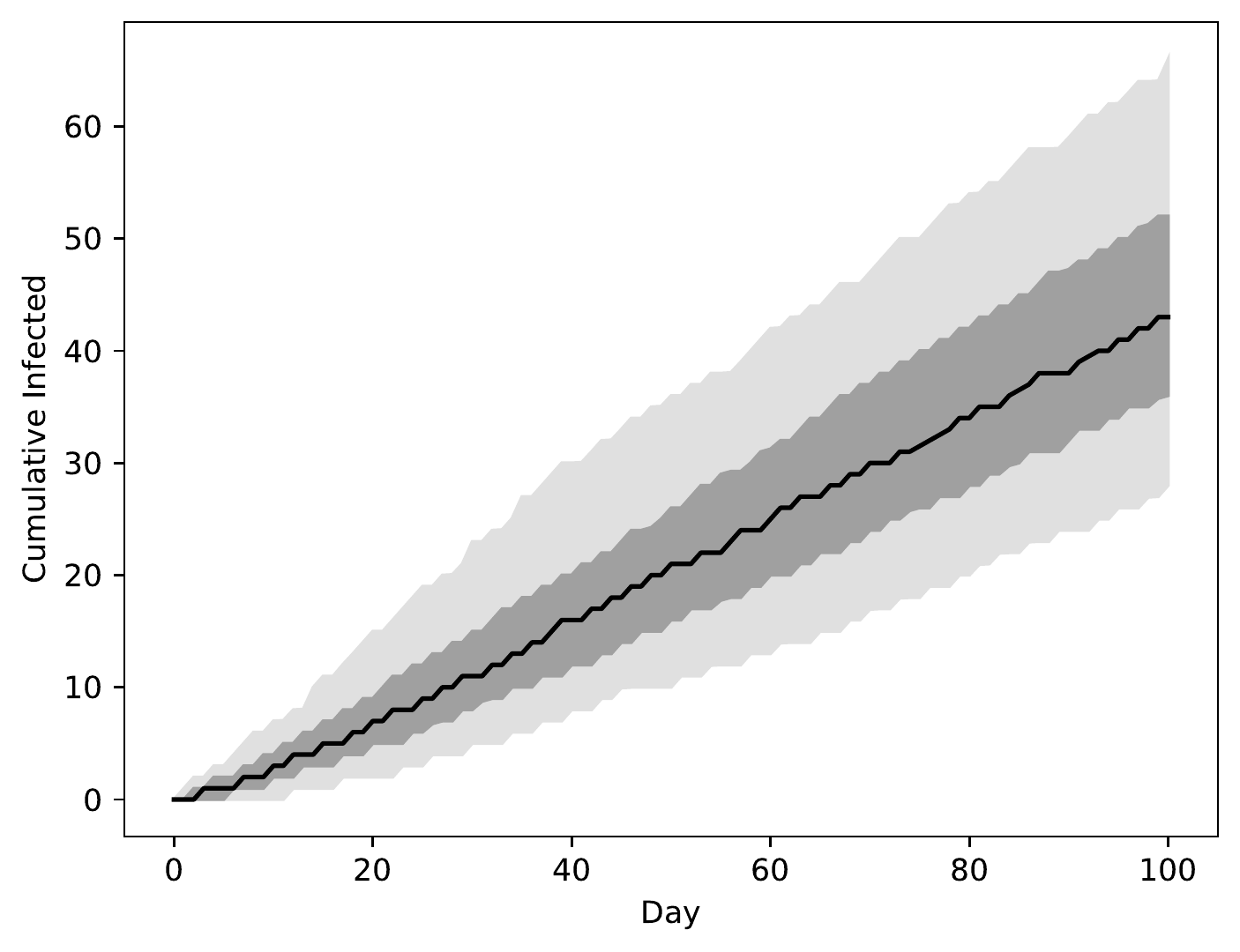}
\caption{Standard Intervention}
\end{subfigure} \\
 \begin{subfigure}[ht]{.5\textwidth}
\centering
\includegraphics[width=\textwidth]{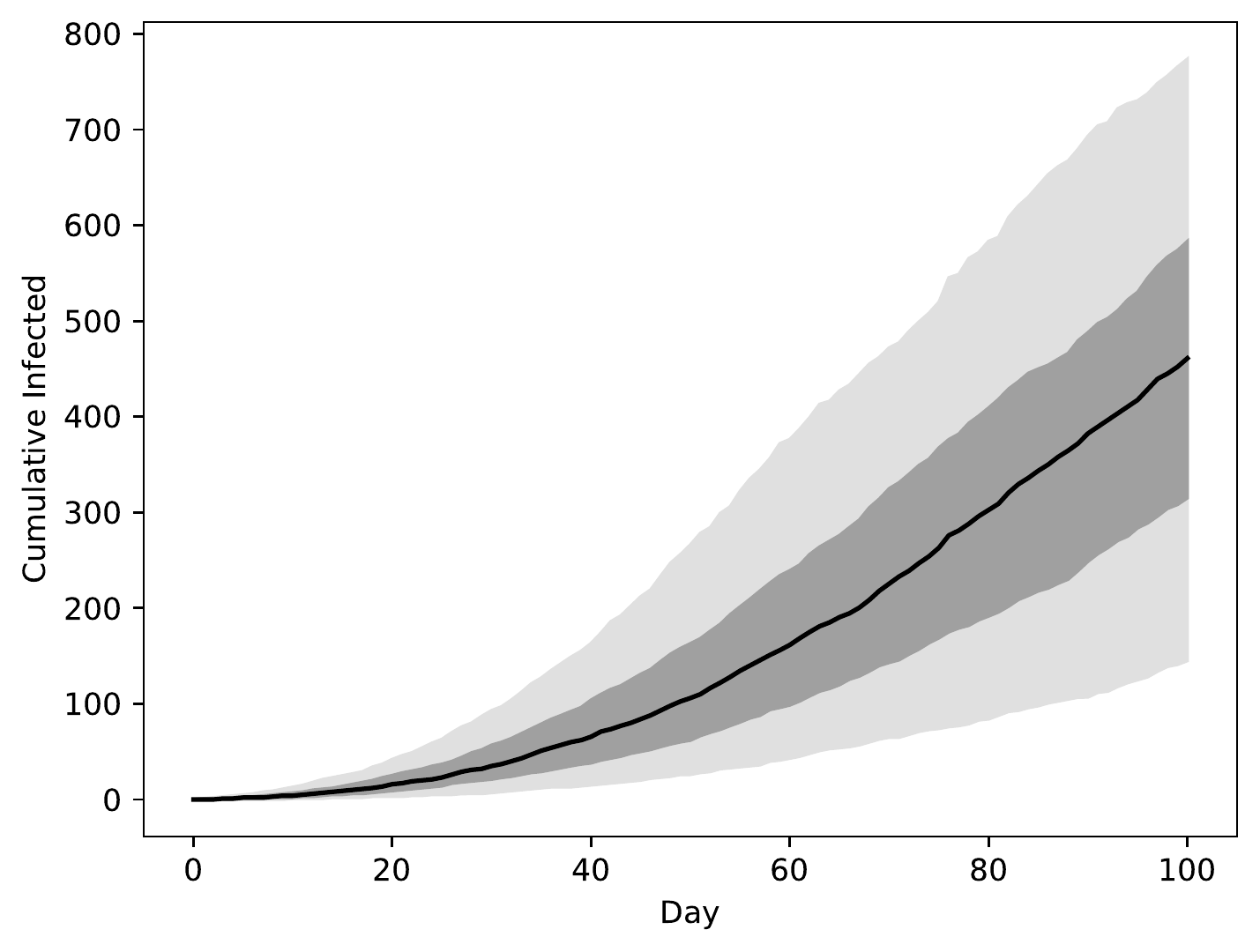}
\caption{Standard Intervention with High Social Contact}
\end{subfigure} \\
\end{center}
\caption{Cumulative infections over a 100-day semester. The black curve is the median daily value for 500 independent runs of the simulation and the dark and light regions indicate quantiles around the median containing 50\% and 90\% of outcomes, respectively.  Under no intervention the median cumulative number of new infections reaches 20,126 (falling between 20,026 and 20,212 in 90\% of simulations), or 89.4\% of the total campus population.}
\label{cumulative_nointervention}
\end{figure}

Although implementing the standard intervention is costly, it is also crucial for controlling disease outbreaks.  In the absence of any intervention, all scenarios end with effectively all susceptible community members developing COVID-19 by the end of the semester, with peak infection rates reached between 20 and 40 days into the semester (Figure \ref{cumulative_nointervention}).  In contrast, the standard intervention avoids the epidemic tipping point altogether (i.e., $R_0$ remains well below $1$) and keeps cumulative infections below 66 in more than 95\% of simulations.\footnote{Regular flatter portions of the graph indicate weekends: new infections are largest during the week when classes are in session. We note that others have observed that weekends play a critical social distancing role in reducing influenza spread \cite{cooley}.}  As we discuss later, the results are sensitive to the rates of social contact among students: increasing expected broad social contacts from 2 per day (as in panel (b)) to 17 per day (as in panel (c)) increases the median cumulative number of infections from 44 to 461 even with the full standard intervention.

\begin{figure}[h]
\begin{center}
\begin{subfigure}[h]{.5\textwidth}
\centering
\includegraphics[width=\textwidth]{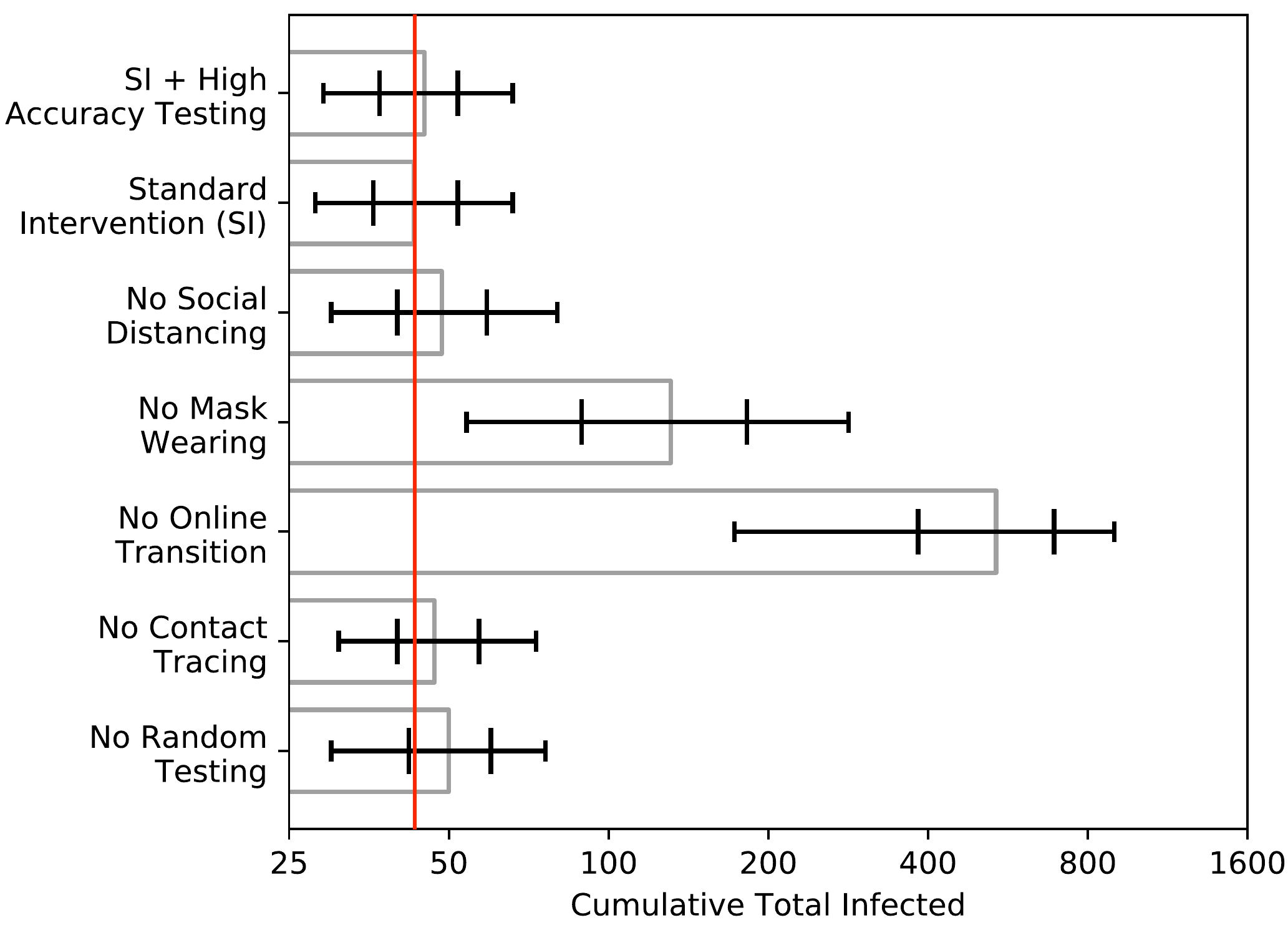}
  \caption{Total Infections}
\end{subfigure}%
  \begin{subfigure}[h]{.5\textwidth}
\centering
\includegraphics[width=\textwidth]{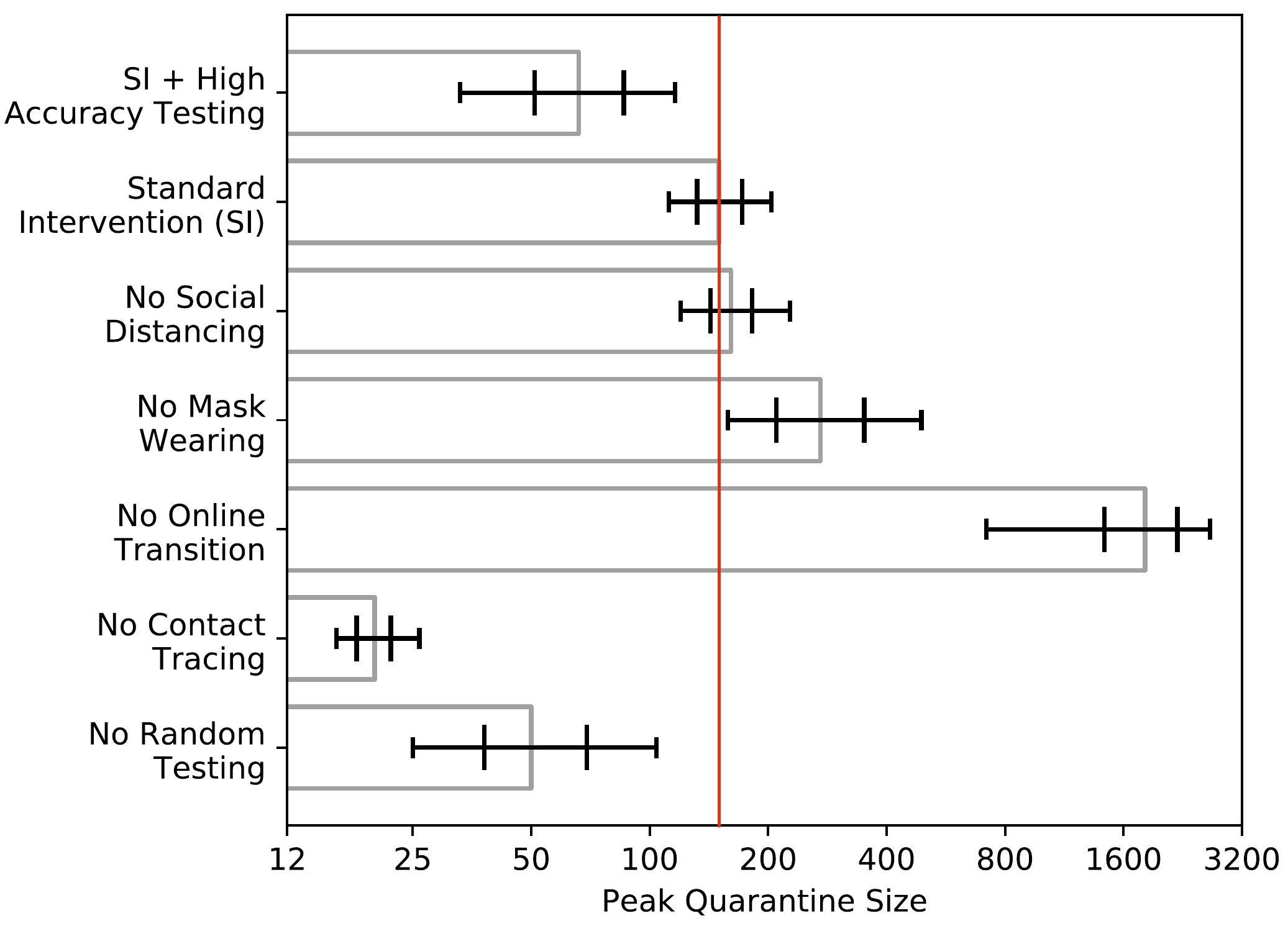}
\caption{Peak Quarantine Population}
\end{subfigure} \\
\end{center}
\caption{Median number of total community members infected over the course of the semester and peak number of students in quarantine by intervention bundle, plotted on a log scale.  Bars indicate quantiles around the median containing 50\% and 90\% of outcomes, respectively.}
\label{measure_comparison}
\end{figure}

Which measures in the standard intervention are the most important?  To control infections, the online transition for classes with more than 30 students is very effective: allowing in-person meetings of large classes increases infections from 43 to 538 in the median simulation run (Figure \ref{measure_comparison}).  Requiring masks is moderately important: not requiring masks increases median infections to 131.  Random testing and contact tracing have the lowest individual impacts, and removing either of these measures (while keeping all others) increases median infections to 50 and 47, respectively.\footnote{We note, however, that although random testing has only a small direct effect on reducing infections it may be essential for monitoring and responding to the prevalence of the disease on campus.}  The effect of moving classes into larger spaces (social distancing) is also small; failing to capitalize on this opportunity raises the median infection total to $48.5$.\footnote{Infection and quarantine rates for additional intervention bundles are reported in the robustness checks in Appendex Tables \ref{appendix:contact1} and \ref{appendix:contact2}.}

The standard intervention results in a median peak quarantine level of 150.  In-person large class meetings drastically increase this number both by increasing the number of infections and increasing the number of students exposed to each positive case: \textit{1,815 students} are in quarantine at peak.  Removing contact tracing and random testing all reduce this number (to 20 and 50)  at the cost of increasing the cumulative number of infections.  Not requiring masks nearly doubles the peak number of quarantined students to 272.
To emphasize the challenge these numbers represent, let us also note that in the standard intervention, the median number of \textit{unique} individuals quarantined at some point during the semester is 602, which means that even in this full-strength mitigation scenario, it often happens that 3\% of all students spend some fraction of the semester in quarantine.

As noted in Section \ref{dynamics}, we find that the false positive rate has a very large effect on the peak quarantine size. Table \ref{fpr2} in the Appendix, for example, shows that a FPR of $2\%$ can increase the median peak quarantine size from $150$ to 1,443. Given the uncertainty around what constitutes a realistic FPR, we also simulated a ``high accuracy'' testing regime in which the FPR is set to $10^{-6}$ and the false negative rate is set to $6\%$. This corresponds roughly to a policy of mandatory retesting of every positive result and a presumption of negativity when the second test is negative. Even with the increased false negative rate, the change to high accuracy testing has only a minor effect on total infections (the median is $45$ instead of $43$). The main difference is a more than twofold reduction in the number of students in quarantine: median peak is lowered from 150 to 66. The effect on the total number of individuals quarantined is even more dramatic: the median number of unique individuals experiencing quarantine drops from 602 in the standard intervention to 178 with high accuracy testing.

We also examine the sensitivity of these results to increasing the rates of different types of social interactions. We increased interaction rates for academic interactions (in and around the classroom), residential, and broad social contacts to simulate an additional 5, 10, or 15 contacts per day.  (In the case of academic contacts, we note that the standard intervention leads to only some fraction of these additional contacts actually occurring). Table \ref{newcontact} summarizes the new levels of cumulative total infected and peak quarantine size under these different scenarios.

\begin{table}[h] 
\begin{center}
\begin{tabular}{|l||c|c|c|}
\multicolumn{4}{c}{Cumulative Total Infected} \\
\hline
& \multicolumn{3}{c|}{Extra Contacts/Day} \\
\hline
& +5 & +10 & +15 \\ \hline \hline
Academic & 50 & 60 & 71.5 \\ \hline
Residential & 89.5 & 163.5 & 279 \\ \hline
Broad Social & 72 & 162 & 461 \\ \hline
\end{tabular}
\ \ \ \
\begin{tabular}{|l||c|c|c|}
\multicolumn{4}{c}{Peak Quarantine Size} \\
\hline
& \multicolumn{3}{c|}{Extra Contacts/Day} \\
\hline
& +5 & +10 & +15 \\ \hline \hline
Academic & 168 & 197 & 227 \\ \hline
Residential & 334 & 621 & 1059.5 \\ \hline
Broad Social & 408.5 & 964 & 2587.5 \\ \hline
\end{tabular}
\end{center}
\caption{Effect of increased rates of contact on cumulative total infected (left) and peak quarantine size (right).}
\label{newcontact}
\end{table}

Our model is relatively insensitive to additional academic contacts; an additional 15 academic contacts per day almost doubles the overall rate of academic contact but does not double the total number of infected individuals or the peak quarantine size. On the other hand, an increase of 5 additional residential contacts raises the total number of infections to 89.5, which is more than double the default level of 43. Peak quarantine size is also more than doubled. Increasing broad social contacts by 5 per day has a slightly weaker effect on total infections than does the corresponding increase for residential contacts, but peak quarantine size increases to 476, which is $272\%$ of the default level. Thus there is a clear difference in our model between the risk involved with additional academic versus non-academic contact, and it is the latter which has stronger undesirable consequences. Between residential and broad social contacts, the comparison is more difficult, but as the number grows larger, broad social contacts become clearly more problematic than residential contacts. This is likely due to the fact that residential contacts are far more compartmentalized than broad social contacts, which places stronger limits on the overall rate of transmission that can occur exclusively within the residential contact network. Because residential contacts are a closer form of contact than broad social contact, it further seems relatively unlikely that real-world students would be able to form very close residential contacts with 10 or more individuals, while it would be rather easy to come into regular broad social contact with much larger numbers of people through extracurricular activities and social events.

\subsection{Moving Large Classes Online}
A key component of managing both infection rates and quarantine peak is to move the largest classes online.  Our baseline standard intervention scenario moves all classes with more than 30 students online; this represents approximately 20\% of the classes in our modeled university (Figure \ref{sizes}).  This intervention is highly effective and presents important tradeoffs.  Teaching effectively online is more costly in terms of course preparation and potentially less appealing to students.  Limiting large classes also drastically reduces infection rates by reducing the number of students exposed to an infected individual and reduces quarantine rates by reducing the number of students identified for quarantine via contact tracing in response to a positive test.  Large classes are also more dangerous for instructional staff.  Moving large classes online has the added benefit of facilitating social distancing measures in existing classes by making larger capacity classrooms available for smaller classes.

\begin{figure}[h]
\begin{center}
\begin{subfigure}[h]{.5\textwidth}
\centering
\includegraphics[width=\textwidth]{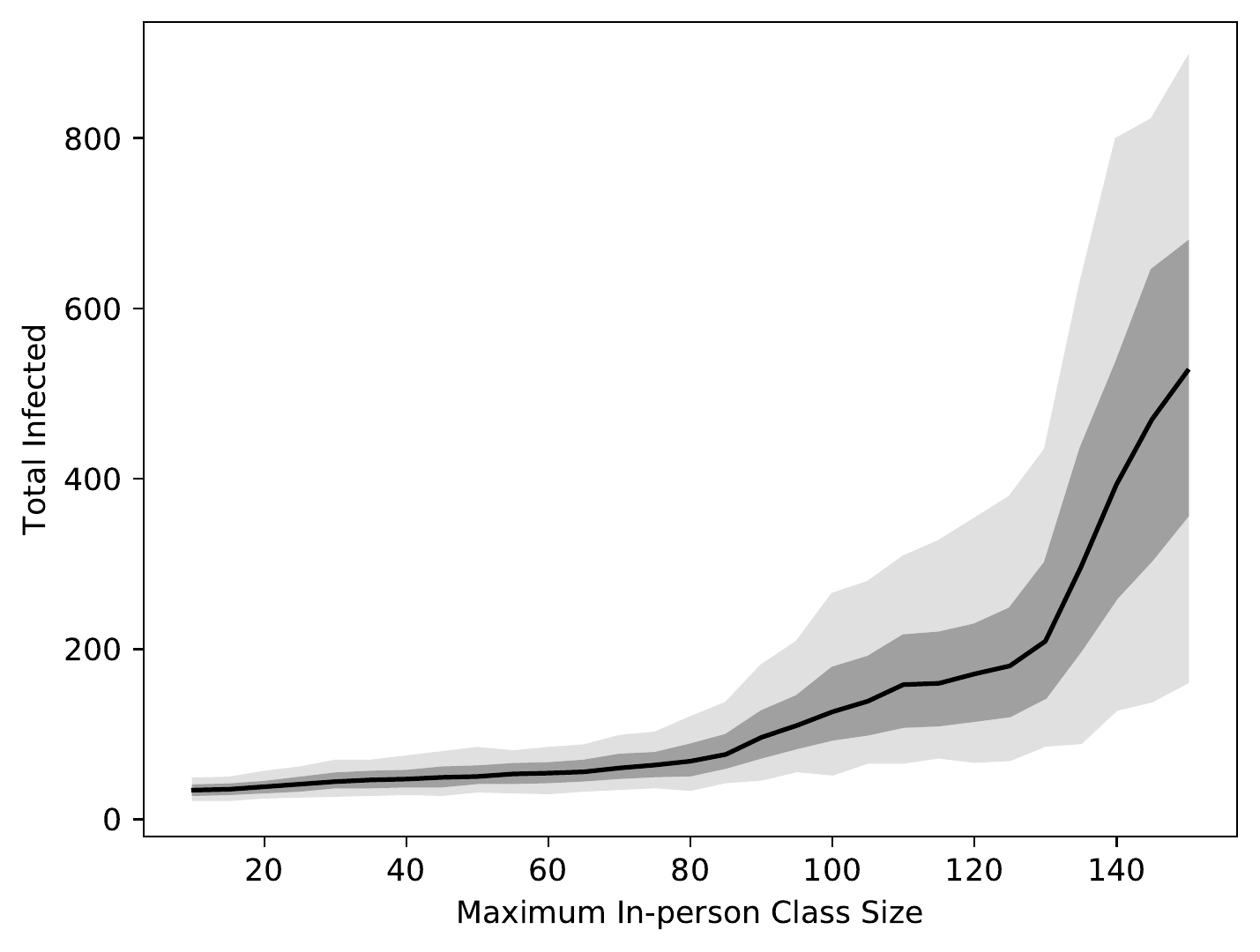}
  \caption{Total Infections}
\end{subfigure}%
  \begin{subfigure}[h]{.5\textwidth}
\centering
\includegraphics[width=\textwidth]{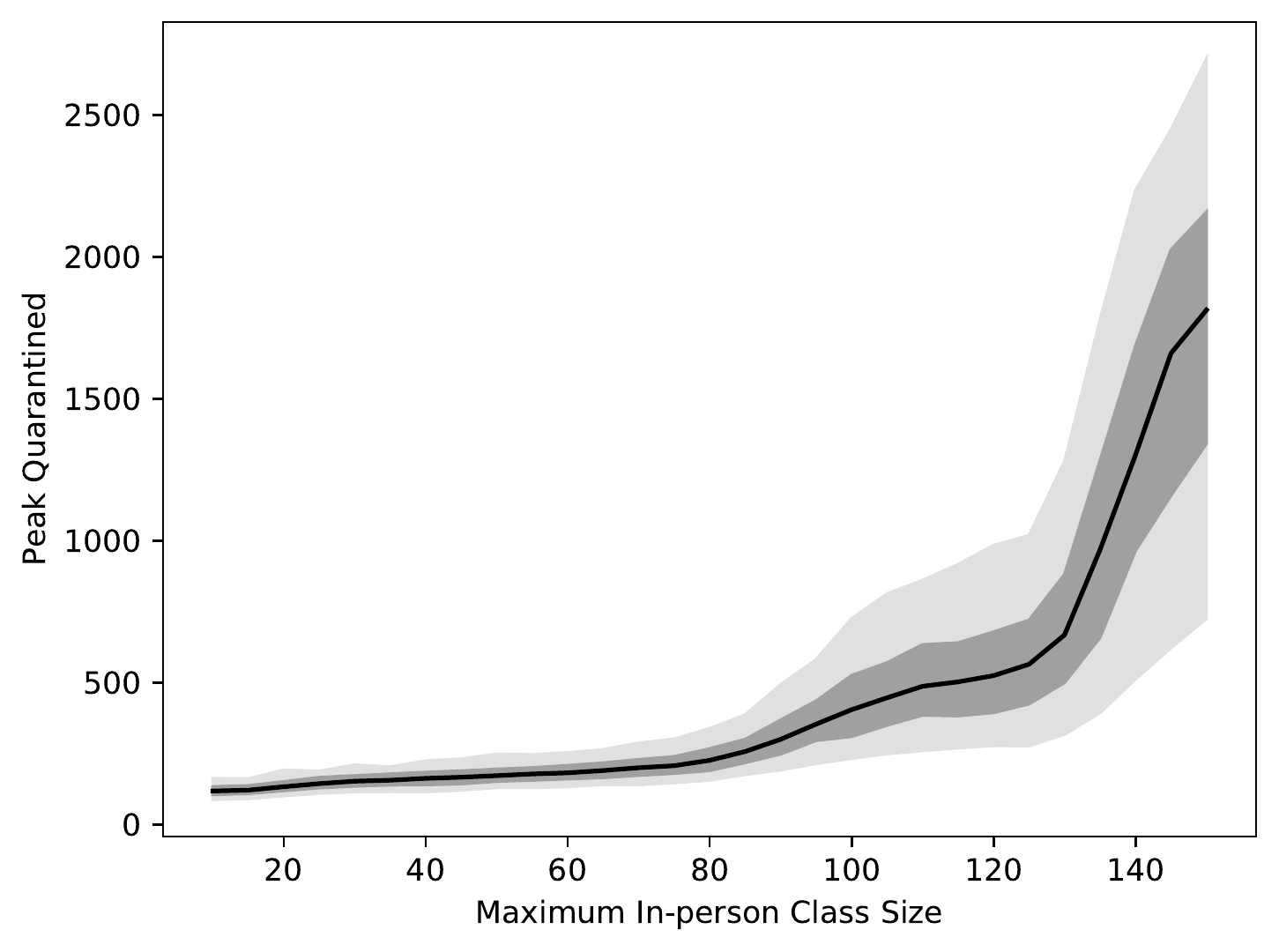}
\caption{Peak Quarantine Population}
\end{subfigure} \\
\end{center}
\caption{Total infections and peak quarantine population as a function of the online class size threshold.  The black curve is the median value for 500 independent runs of the simulation and the dark and light regions indicate the 50\% and 90\% quantiles, respectively.}
\label{size_sweep}
\end{figure}

How sensitive are our outcomes to the threshold for moving to online instruction?  
Figure \ref{size_sweep} shows the response of both the infection count and peak quarantine to changes in the in-person class size cap.  These figures show that eliminating the largest classes is crucially important: allowing all classes to meet in person increases the median number of infections from 43 to 527 and the median peak quarantine from 150 to 1,813.  In 5\% of simulations, the peak number of students in quarantine reaches 2,697 and 893 people are infected by the end of the semester.  The responses to class size are very nonlinear: relaxing the cap at 30 leads to only mild growth in infection and quarantine, while increasing beyond 100 leads to larger growth, and increasing beyond 120 rapidly expands the magnitude of the outbreak. The exact threshold at which this rapid escalation occurs is, of course, a function of the other interventions in place (3\% daily random testing, contact tracing and quarantine, and mask wearing) and should not be expected to persist in all or even most real-world environments. However, there does appear to be a general principle at work here, which is that although the connectivity of the academic network graph is still high, the number of transmission opportunities is greatly diminished: with no online transition, students are classmates with an average of $1.2\%$ of the student body, and when classes of 30 and greater transition online, the proportion drops to only $0.2\%$. This roughly means that more than 80\% of the edges in the graph connecting students result from the largest 20\% of classes. Such a drop does not in and of itself guarantee successful containment, but in our model combines with the other interventions to result in a decrease of the median reproduction number from $1.125$ in the uncapped case (which is technically uncontrolled, but close enough to $1$ that exponential growth is difficult to detect on the span of $100$ days) down to $0.44$, which is safely outside the range of uncontrolled exponential growth.

\begin{figure}[h!]
  \begin{center}
  \includegraphics[width=.5\textwidth]{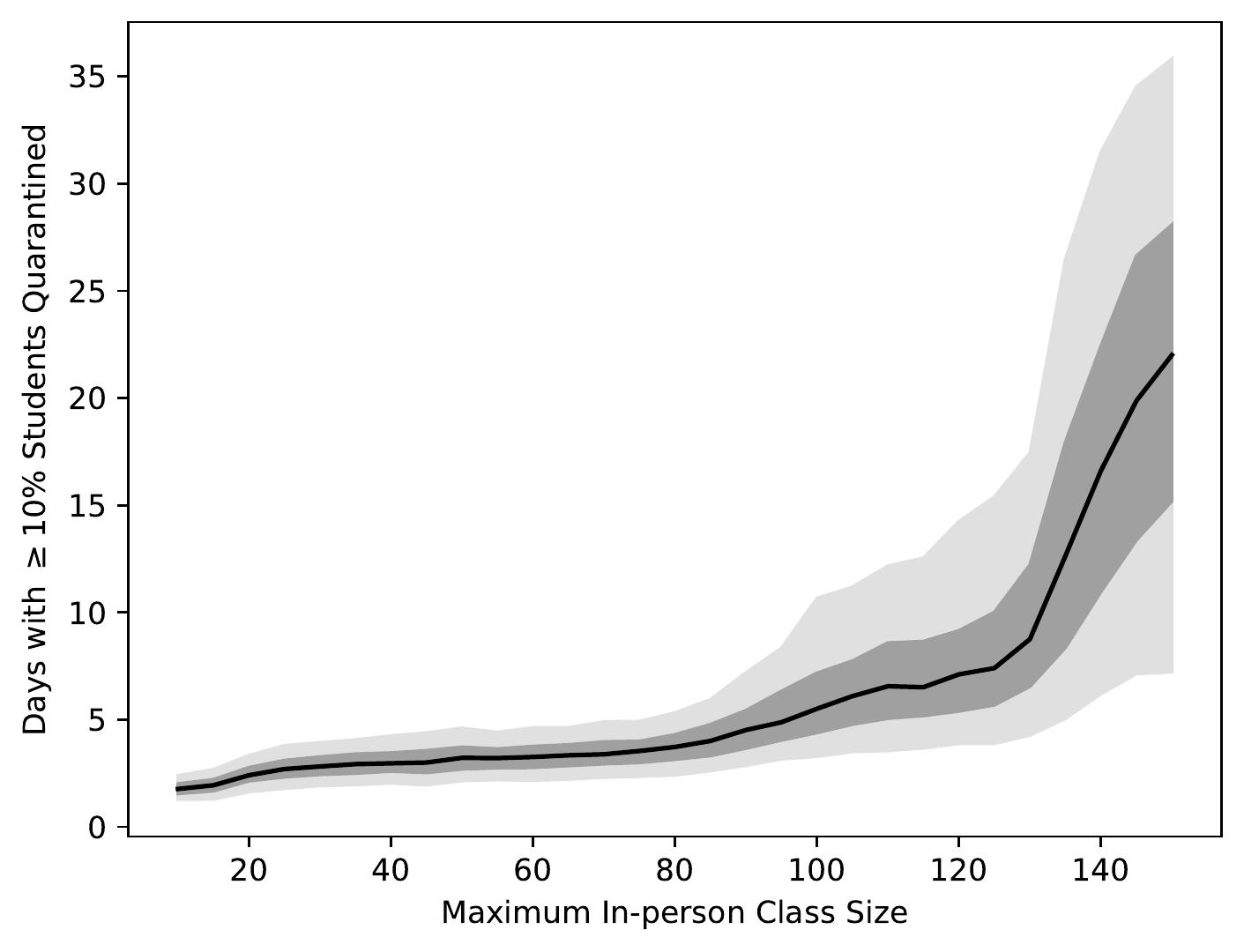}
  \end{center}
  \caption{Number of days that an average class experiences $\geq 10\%$ of students in quarantine as a function of the online class size threshold.  The black curve is the median value for 500 independent runs of the simulation and the dark and light regions indicate the 50\% and 90\% quantiles, respectively.}
  \label{absent}
\end{figure}

The class size cap also has a significant impact on the viability of courses as the semester progresses.  Figure \ref{absent} plots the number of days 
in which more than 10\% of students in an average class are quarantined as a function of the online class size threshold.  This reflects a heightened likelihood that a substantial number of classes have a practically meaningful number of students absent due to quarantine.  This follows the pattern of the number of students in quarantine: there is prevalent absenteeism for about $5.5$ days on average when classes are capped at 100 students, with this increasing to about $10.5$ days for the top 5\% most extreme simulations.  Capping class size at 150 leads to an expected 22 days of significant absenteeism, with the top 5\% of simulation runs resulting in more than five weeks of widespread absence.  This presents a significant challenge for large, in-person classes: holding large $100+$ student classes in person does not resolve the problem that those classes spend weeks of the semester with at least $10$ students in quarantine and in need of remote instruction. Planning for this by offering large classes online has the added benefit of decreasing absenteeism in smaller classes as well.

Finally, while we have chosen our parameter values (number of daily contacts, $R_0$, external infection rates, FPR and FNR) to be as realistic as possible, it is important to examine how our results vary with these inputs.  These results are discussed in more detail in the Appendix, but broadly indicate that the standard intervention is relatively effective across a range of parameter values.
Infections and quarantine remain manageable (less than 200 infections and 300 in quarantine) under the standard intervention as long as daily contacts are less than 31.5 (baseline of 19) and the expected number of external infections is less than 1.0 or 2.0 per day (baseline of 0.25) (Tables \ref{appendix:contact1}, \ref{appendix:contact2}, \ref{outside1}, and \ref{outside2}).  These external infection rates reflect a scenario where the spread of coronavirus regionally is well under control: a local outbreak of the virus will be reflected in highly elevated infection and quarantine numbers on campus.
The standard intervention bundle is extremely robust to all reasonable values of $R_0$: predicted infections and number quarantined reach a peak of 73 and 193 respectively when $R_0$ is raised from 3.8 to 5.8.  Outcomes under less rigorous intervention bundles are much more sensitive to increases in these parameters.  The total number of infected is largely insensitive to changes in testing accuracy (both the FPR and the FNR), while quarantine grows very quickly with the FPR (but not the FNR).

\section{Conclusion}
\label{conclusion}

The control of the spread of the novel coronavirus on campus is a critical challenge for colleges and universities planning to open for in-person instruction in the fall.  Our agent-based model of the spread and containment of this disease on campus is an effort to be as realistic as possible: engaging the current state of the epidemiological understanding of this illness as well as capturing some of the idiosyncrasies of the residential university setting.  We are able to model behavior that is highly structured and fully individualized using a relatively manageable number of agents.  This allows our model to more accurately capture disease dynamics than a traditional compartmental model.  We find important implications for the impact of different control measures on both the spread of the disease and on the disruption of academic life.  Our results suggest that it is important to have a robust portfolio of interventions, and that universities should be keenly focused on the specificity of their testing regime.  Moving the largest classes online is an effective tool in the arsenal to stop the spread of the disease and to minimize the disruption caused by quarantining potentially exposed students.  We also find significant differences between the risks involved with academic and other types of regular contact, with spread through broad social contact representing being more dangerous than residential contact in the extreme of many additional contacts.  

There are several important limitations of our analysis.  First, we have not included seasonal effects in transmission patterns or the arrival of infections from off-campus.  These effects may require additional control measures and have motivated several universities to shorten their semester to avoid extending into December.  To gain some very basic understanding of the potential magnitude of the challenge this may present, we have included results in Tables \ref{outside1} and \ref{outside2} in the Appendix which suggest that the total number of those infected and the peak quarantine size depend linearly on the rate of infections from outside the university. There may also be additional unmodeled heterogeneity in the intensity of contact between students in different types of class settings: interactions in large lectures, small seminars, and labs may yield different transmission rates between students.  Some students may also opt to continue their instruction remotely (e.g., due to travel restrictions), which will further decrease on-campus transmission for a given enrollment. Finally, we have not accounted for the fact that compliance with mask-wearing policies will likely be diminished in residential settings and social settings; our modeling leads us to believe that that this will lead to modest increases in infection and quarantine rates.

Although we include instructors in this model, their off-campus behavior and infection risk is largely unmodeled. In the standard intervention, we find that a median of only 5 instructors are infected during the semester, so the effect of additional within-university interactions (between faculty and staff or administration, for example) may ultimately be a more significant risk to faculty than direct faculty-student interactions.  Faculty and staff may also be exposed to the disease through their social and residential contacts: in an urban setting, for example, public transit may be a particularly significant risk to faculty and staff and may indirectly lead to more student infections as well.

Finally, this model has limited applicability to other types of university settings, such as partial-residential and non-residential campuses.  Small colleges may be importantly different than the university modeled here in that large classes are less common (and smaller) and that social networks may differ in ways that affect disease spread. Liberal arts colleges seem to present an especially difficult challenge since daily academic contacts would be much fewer (in a way that should be relatively easy to model) but of a sort more likely to transmit infection (which would be a more difficult feature to reasonably quantify or calibrate without additional data).  Nonetheless, many of the main observations are likely to generalize well: increasing testing accuracy, limiting large classes, minimizing non-residential social contacts and preparing to accommodate unpredictable absences are likely to be useful across many contexts.

\FloatBarrier

\bibliographystyle{siam}
 \newcommand{\noop}[1]{}

\newpage

\section{Appendix}

\subsection{Additional Model Details}

\subsubsection{University Structural Modeling}
\label{appendix:enroll}
As described in Section \ref{contactpatterns}, our goal is to generate simulated course rosters which yield a realistic distribution of class sizes as summarized in Figure \ref{sizes}. The process used to accomplish this task is as follows:
\begin{itemize}
\item From fixed histogram data, we generate a static probability density function for class size. Classes are assumed to be arranged from largest to smallest, and this ordering is also regarded as an ordering from less advanced to more advanced. To accelerate the process of generating simulated schedules, we cluster classes into groups of 5 (by order) to give a total of 750 different groups.
\item Students randomly and independently choose 4-5 different course clusters. Once the clusters are selected, courses are assigned by choosing specific classes within each cluster uniformly at random. This gives a computationally efficient way to produce the desired marginal distribution of class sizes. Because the distribution is not uniform, we found that other methods (e.g., drawing from a full list of 3750 and repeating until all 4-5 selections were distinct) resulted in changes to the marginal distribution which tended to somewhat overrepresent large courses. The advantage of our approach is that the final marginal distribution of class sizes will simply be proportional to the selection probabilities.
\item To account for the differing enrollment behaviors of students in different courses, we construct different probability densities for each cohort. To keep the marginal distribution fixed, we construct custom distributions as follows. Let $X_0,\ldots,X_7$ be i.i.d. random variables with a fixed distribution on the set $\{1,\ldots,750\}$. From these variables, we let $Y_0,\ldots,Y_7$ be the values of $X_0,\ldots,X_7$ sorted from smallest to largest, i.e., $Y_0 \leq Y_1 \leq \cdots \leq Y_7$ and $Y_0 = X_{i_0},\ldots,Y_7 = X_{i_7}$ for $i_0,\ldots,i_7$ distinct. When selecting course groups, students in cohort $k$ choose using the same distribution as $Y_k$. This yields, for example, a very small but nonzero probability that a course will be simultaneously taken by a student in cohort 0 (freshmen) and another in cohort 7 (advanced graduate students).
\item To group students into close contacts/friends, we simply order students in each class of at least 5 by the order in which they enrolled and form groups of minimal size meeting or exceeding the desired target size (the square root of the overall size). This yields at most one group in each class which is significantly smaller than the target and no groups which are significantly larger. Because each student's cohort is chosen uniformly at random at the time of enrollment and students choose their courses independently, the order of students within a single course is effectively uniformly random. With this method, there is the possibility for correlation of grouping across classes when students have more than one class in common, which desirable in a model of close social contacts.
\end{itemize}

\subsubsection{Contact Processes}
\label{appendix:contacts}
As mentioned earlier, our model has two basic categories of contact which may occur; the first, Poisson contact, is generated via a Poisson point process. For two individuals, $a$ and $b$, the process may generate a contact which allows $a$ to transmit to $b$ (which we denote $a \rightarrow b$) or vice versa ($b \rightarrow a$). Poisson point processes also sometimes generate contacts with multiplicity greater than $1$, which we interpret to mean that multiple instances of the same sort of contact occur on the same day; each instance is regarded independently of any others from the perspective of determining whether a susceptible individual becomes infected. To be clear, any two contacts are generated independently of one another, so that $a \rightarrow b$ can occur whether or not $b \rightarrow a$ occurs. In most cases, each of these two options is as likely as the other, but we do allow for one exception: within the classroom, we assume that instructors are more likely to transmit to a student than the student is likely to transmit to them. This is meant to reflect the situation that a professor actively lecturing for an hour is likely to shed more virus particles than a student is while listening to the lecture.

The rate for a particular contact $a \rightarrow b$ on any given day is a sum of a number of individual terms which depend on the activities that $a$ and $b$ have in common. Below we list all the possible contributions. In each case, the rates described below are all summed to determine the overall rate. If a category does not apply to a particular pair $a,b$, the corresponding contribution to the rate for that day is simply zero. Note that residential contacts are an exception and do not contribute to the Poisson rate; instead, residential contacts are guaranteed to occur every day and are counted with multiplicity $1$ in addition to all the contacts generated by the point process.
\begin{itemize}
\item Poisson Contacts: Academic contacts \\
There are five subcategories of interactions which are included under the broad heading of Poisson contacts which are collectively referred to as ``academic contacts.''
\begin{enumerate}
\item Close Academic Contacts (average 4/day by default with no interventions applied)
\begin{itemize}
\item Modeled Behaviors: students regularly sitting near one another in class, occasionally meeting outside of class to work on class activities, making friends in class; instructors having departmental neighbors or friends
\item Mechanics: Classes and recitations of size $N \geq 5$ are subdivided into groups of size $\sqrt{N} \pm 1$ (all but one group must lie in this range). Students in the same group receive a fixed contribution to the Poisson rate for their contacts (both $a \rightarrow b$ and $b \rightarrow a$ occur at the same rate) on ``active'' days when the corresponding class or recitation is in session. On inactive days, the contribution to the rate is $1/4$ the active rate. The rate is never adjusted for social distancing.

Likewise, departments of size $N$ are subdivided into groups of size $\sqrt{N}$. Instructors in the same group experience the active rate on Monday through Friday and experience $0$ contact rate for this contact on weekends.

By default, it is normalized to produce an average of $4$ contacts per day for an average individual under normal circumstances (i.e., $a$ experiences approximately $4$ contacts of the form $a \rightarrow b$ or $b \rightarrow a$ for other individuals $b$). In expectation, half of those contacts involving persons $a$ and $b$ will be opportunities for $a$ to transmit to $b$ and vice-versa. Contacts from a Poisson point process occur with multiplicity, so it is also possible for several instances of $a \rightarrow b$ to occur in a particular day, and each is treated independently (so that multiple contacts increase the probability of transmission). For similar reasons, it is also possible for potential transmissions to occur in either direction $a \rightarrow b$ or $b \rightarrow a$ on the same day.
\end{itemize}
\item Classroom Contacts (average 4/day when combined with Department Professional Contacts below)
\begin{itemize}
\item Modeled Behaviors: Classroom interactions
\item Mechanics: Each class meets on a recurring schedule which is either MWF (40\% of classes), TTh (40\%), or MW (20\%). Recitations meet one weekday per week on a day in which the corresponding section does not meet. On days when the class or recitation is in session, students receive a fixed contribution to their contact rate for each class that they have in common. For individuals $a,b$ in the classroom, the rate of $a \rightarrow b$ contacts is proportional to a product $I_a S_b$, where $I_a = S_a = 1$ when $a$ is a student, $I_a = 4, S_a = 2$ if $a$ is an assistant attending the section they are assigned to, and $I_a = 10, S_a = 5$ if $a$ is the instructor of the class or is the assistant in a recitation.

The rate of contact is also affected by social distancing when available. It is assumed that all classes, recitations, etc., which are over the online transition threshold vacate a physical room which can then be claimed for use by a smaller class or recitation. New rooms are assigned beginning with the largest active classes first and moving to smaller ones. If the class whose space was vacated (the ``vacated class'') had at least $20$ students and was strictly more than $50\%$ larger than the active class moving up, the rate of contact in the active class is multplied by the factor
\[ C_{SD} = \min \left\{ 1, \frac{ \max \{ \text{class size}, 10 \}}{ \text{vacated class size}} \right\}. \]
When social distancing is active, an average value of this constant is computed by weighting classes according to the square of their size. This is called the ``crowd reduction factor'' and is used in other instances of social distancing listed below.

When social distancing is enabled, the contact rate in every classroom is also multiplied by the proportion of nonquarantined students each day contacts are constructed. This feature models a small amount of additional distancing which is possible when attendance rates are lower than 100\%.
\end{itemize}

\item Department Professional Contacts
\begin{itemize}
\item Modeled Behaviors: Instructors interacting in common departmental areas like mailrooms, department meetings
\item Mechanics: On Mondays through Fridays, all instructors in the same department experience a fixed rate of contact (equal for $a \rightarrow b$ and $b \rightarrow a$).  The rate is fixed to be $8$ times the rate of student-to-student contact in a classroom without social distancing.
\end{itemize}

\item Department Environmental Contacts (average 4/day by default with no interventions applied)
\begin{itemize}
\item Modeled Behaviors: sharing elevators, walking through common hallways, touching surfaces or doorhandles, eating food at nearby food trucks, etc.
\item Mechanics: Every day that an individual (instructor, student, or assistant) travels to any given department, that individual belongs to a pool associated to that department. If an individual has multiple appointments in the department, they occur in the pool with a corresponding multiplicity. Poisson contacts are drawn between pairs of individuals in the pool at a rate which is proportional to the product of their multiplicities. When social distancing is active, the rate is also multiplied by the crowd reduction factor from above.

These contacts are assumed to be non-traceable, which means that they are not included in any contact tracing activities. They do not occur on weekends.
\end{itemize}
\item Broad Environmental Contacts (average 4/day by default with no interventions applied)
\begin{itemize}
\item Modeled Behaviors: interactions on the street going to or from class, contact in local businesses or restaurants, infrequent professional interactions of individuals across departments, etc.
\item Mechanics: Every day that an individual has any physical class, they belong to a pool of individuals active that day. As with department environmental contacts, individuals occur with multiplicity if they have multiple physical destinations that day. The rate of contact is proportional to the product of multiplicities, and also to the crowd reduction factor when social distancing is available.  These contacts are non-traceable and do not occur on weekends.
\end{itemize}
\end{enumerate}
\item Poisson Contacts:  Broad Social Contacts (average 2/day by default with no interventions applied)
\begin{itemize}
\item Modeled Behaviors: Dining halls or restaurants, social events, extracurricular activities
\item Mechanics: Any two individuals experience broad social contact on a given day with a rate which is independent of the individual. The rate of broad social contact is twice as high on the weekends as it is on weekdays. 
\end{itemize}
\item Residential Contacts (by default: each student linked to an average of one other)
\begin{itemize}
\item Modeled Behaviors: Students living in residential colleges
\item Mechanics: In each cohort of students, individuals are randomly ordered. Each individual is linked with a random number of predecessors in the cohort, drawn from an exponential distribution whose mean is $1/2$ the desired number of neighbors. Links are considered symmetric, so that each student is also linked to another $1/2$ the desired number of neighbors by symmetry.  Note that links are specifically not transitive, so that it is possible to find long chains for potential transmission when the number of neighbors is moderately large.

When students $a$ and $b$ are linked residentially, the contacts $a \rightarrow b$ and $b \rightarrow a$ are both guaranteed to occur at least once every day (they can occur more than once if these contacts also arise from other means). In this sense, a residential contact is a stronger form of contact than either acadmeic or broad social contacts, more akin to a roommate or a very close neighbor than to casual friendships (which are closer to broad social contacts). 
\end{itemize}
\end{itemize}

\subsubsection{Discretized Gamma Distribution}
We use a discretized Gamma distribution to generate both the incubation period and daily infectiousness for each infected individual. The Gamma distribution is widely used in connection with COVID modeling \cite{imperial1} for these parameters; we formulate a discretized version to ensure accuracy in our discrete-time simulation.

We let $\Gamma_d(k,\theta)$ denote our discretized Gamma distribution of shape $k$ (a positive integer) and scale $\theta > 1$. The variable takes values in the nonnegative integers and has PDF \label{gamma}
\[ \mathrm{P}( \Gamma_d(k,\theta) = n) := \frac{n \cdots (n+k-2)}{(k-1)! \theta^k} ( 1 - \theta^{-1})^{n-1}. \]
To see that the probabilities sum to $1$, one can simply differentiate the power series expansion for $(1-x)^{-1}$ a total of $k-1$ times and perform a reindexing. Specifically,
\begin{align*}
 (k-1)! \theta^k & = (k-1)! (1- (1 - \theta^{-1}))^{-k} = \left. \frac{d^{k-1}}{d x^{k-1}} \right|_{x=1-\theta^{-1}} (1 - x)^{-1} \\
  &  = \left. \frac{d^{k-1}}{d x^{k-1}} \right|_{x=1-\theta^{-1}}  \sum_{n=0}^\infty x^n =  \sum_{n=0}^{\infty} n \cdots (n-k+2) ( 1 - \theta^{-1})^{n-k+1}  \\
  & = \sum_{n=0}^\infty (n+k-2) \cdots n (1-\theta^{-1})^{n-1},
 \end{align*}
 where the last line follows by reindexing the sum by sending $n$ to $n + k-2$. The lower limit of summation does not change because the corresponding terms are zero. The mean $\mu$ and variance $\sigma^2$ of this distribution satisfy
 \[ \mu = 1 + k (\theta - 1), \qquad \sigma^2 = k \theta(\theta-1).\]
 In the former case, one can quickly establish by writing $n = k \theta \frac{n+k-1}{k\theta} + 1-k$ and then rewriting the sum for $\mu$ in terms of the sums just computed for shapes $k+1$ and $k$.  Likewise, in the latter case, one uses the identity
 \[ n^2 = (n +k )(n+k-1) - (2k-1) (n+k-1) + (k-1)^2\]
and relates the resulting series to the distributions of $\Gamma_{d}(k,\theta),\ldots,\Gamma_d(k+2,\theta)$.

\subsection{Robustness Testing}

To map the dependence of our result on various parameter choices, we ran a wide array of 50-simulation batches under various alternate calibrations. In each case, we report the median value (rounded to the nearest whole number) for that calibration under various combinations of interventions. To simplify matters, we adopt the following abbreviations for certain basic combinations:
\begin{itemize}
\item TTQ $=$ mass testing, contact tracing, and quarantine
\item OLSD $=$ transitioning courses of 30 or more students to online-only together with social distancing via moving classes into larger vacated classrooms.
\item SI $=$ TTQ, OLSD, and universal mask-wearing.
\end{itemize}
Tables \ref{appendix:contact1} and \ref{appendix:contact2} show the first collection of such results where the overall rate of daily contact has been varied. For simplicity, each of the various types of contact (academic, residential, and broad social) has been varied proportionally relative to the default configuration (so that $38$ contacts per day would consist of a total of 32 contacts from the close social, classroom, department professional, department environmental, and broad environmental categories, 4 broad social contacts per day, and 2 residential contacts). We opted to end the simulation of very large numbers of contacts in each scenario after a threshold of 3,000 total infected or 5,000 peak quarantine had been reached, as any more extreme scenario than this is clearly to be avoided in reality.  Bolded columns represent the default parameter values of 19 daily contacts, and these scenarios correspond to the ones plotted in Figure \ref{measure_comparison} (but we note that the figure summarizes the results of 500 total simulations).
\begin{table}[H]\caption{Cumulative total infected for various alternate calibrations of contact rates}
\label{appendix:contact1}
\begin{center}\begin{tabular}{|l||C{25.01pt}|C{25.01pt}|C{25.01pt}|C{25.01pt}|C{25.01pt}|C{25.01pt}|C{25.01pt}|}
\hline
 & \multicolumn{7}{c|}{Daily Contacts}\\
\hline
 & $6.3$ & $12.5$ & \textbf{19.0} & $25.3$ & $31.5$ & $50.0$ & $75.0$ \\ \hline \hline
SI & 29 & 37 & \textbf{44} & 65 & 73 & 453 & 1577 \\ \hline
No Mass Testing & 29 & 37 & \textbf{46} & 68 & 125 & 1333 & 2985 \\ \hline
No Distancing & 30 & 37 & \textbf{48} & 65 & 113 & 1135 & 2671 \\ \hline
No TTQ & 31 & 47 & \textbf{83} & 310 & 2066 & 14643 &  \\ \hline
No Masks & 36 & 61 & \textbf{138} & 519 & 1876 & 4056 &  \\ \hline
No OLSD & 42 & 98 & \textbf{510} & 1335 & 2043 &  &  \\ \hline
Only TTQ & 114 & 2883 & \textbf{5398} &  &  &  &  \\ \hline
Only OLSD & 45 & 257 & \textbf{7791} &  &  &  &  \\ \hline
Only Masks & 60 & 3707 &  &  &  &  &  \\ \hline
No Intervention & 3291 &  &  &  &  &  &  \\ \hline
\end{tabular}\end{center}Each cell is the median value (rounded to the nearest whole number) of 50 independent runs of the specified scenario.
The bold column represents the default parameter values of 19 daily contacts.
Empty cells are those scenarios for which some strictly less severe scenario has already reached either 3000 infected or 5000 peak quarantined.
\end{table}
\begin{table}[H]\caption{Peak quarantine size for various alternate calibrations of contact rates}
\label{appendix:contact2}
\begin{center}\begin{tabular}{|l||C{25.01pt}|C{25.01pt}|C{25.01pt}|C{25.01pt}|C{25.01pt}|C{25.01pt}|C{25.01pt}|}
\hline
 & \multicolumn{7}{c|}{Daily Contacts}\\
\hline
 & $6.3$ & $12.5$ & \textbf{19.0} & $25.3$ & $31.5$ & $50.0$ & $75.0$ \\ \hline \hline
SI & 56 & 99 & \textbf{156} & 227 & 281 & 1448 & 5187 \\ \hline
No Mass Testing & 11 & 29 & \textbf{48} & 88 & 167 & 2785 & 6870 \\ \hline
No Distancing & 57 & 106 & \textbf{162} & 244 & 408 & 3426 & 7973 \\ \hline
No TTQ & $*$ & $*$ & \textbf{$*$} & $*$ & $*$ & $*$ &  \\ \hline
No Masks & 61 & 118 & \textbf{275} & 1052 & 3458 & 7661 &  \\ \hline
No OLSD & 96 & 369 & \textbf{1760} & 4220 & 5884 &  &  \\ \hline
Only TTQ & 190 & 4691 & \textbf{8673} &  &  &  &  \\ \hline
Only OLSD & $*$ & $*$ & \textbf{$*$} &  &  &  &  \\ \hline
Only Masks & $*$ & $*$ &  &  &  &  &  \\ \hline
No Intervention & $*$ &  &  &  &  &  &  \\ \hline
\end{tabular}\end{center}
Each cell is the median value (rounded to the nearest whole number) of 50 independent runs of the specified scenario.
The bold column represents the default parameter values of 19 daily contacts.
Empty cells are those scenarios for which some strictly less severe scenario has already reached either 3000 infected or 5000 peak quarantined. A star indicates that the intervention does not include quarantine so peak quarantine is trivially zero.
\end{table}

As can be seen from the table, we note that dropping the OLSD combination of interventions from the standard intervention results in greater numbers of additional infections and peak quarantine size than does dropping the use of masks at any of the simulated contact rates. It is likewise true that dropping OLSD results in more additional infections than does dropping TTQ until the level of 31.5 daily contacts, at which point the effects are comparably severe.

The next series of results demonstrates dependence of our main outcome variables on the chosen values of $R_0$ and the rate of infection from external sources. We find that there is only modest change in the standard intervention as $R_0$ varies over the full interval of reasonable values from $2.8$ to $5.8$.  Similarly, the number of total infections depends in a roughly linear way on the rate of external infections (i.e., there is no indication of strong nonlinear dependence). Results are summarized in Tables \ref{outside1} and \ref{outside2}.
\begin{table}[H]\caption{Cumulative total infected for various alternate values of $R_0$ and rates of infection from outside sources}
\label{outside1}
\begin{center}\begin{tabular}{|l||C{25.01pt}|C{25.01pt}|C{25.01pt}|C{25.01pt}||C{25.01pt}|C{25.01pt}|C{25.01pt}|C{25.01pt}|C{25.01pt}|}
\hline
 & \multicolumn{4}{c||}{$R_0$} & \multicolumn{5}{c|}{External Infections Per Day}\\
\hline
 & $2.8$ & \textbf{3.8} & $4.8$ & $5.8$ & \textbf{0.25} & $0.5$ & $1.0$ & $2.0$ & $4.0$ \\ \hline \hline
SI & 36 & \textbf{44} & 49 & 73 & \textbf{44} & 91 & 181 & 352 & 689 \\ \hline
No Mass Testing & 40 & \textbf{46} & 62 & 87 & \textbf{46} & 101 & 194 & 386 & 764 \\ \hline
No Distancing & 42 & \textbf{48} & 66 & 92 & \textbf{48} & 99 & 197 & 400 & 769 \\ \hline
No TTQ & 50 & \textbf{83} & 180 & 936 & \textbf{83} & 181 & 354 & 686 & 1344 \\ \hline
No Masks & 60 & \textbf{138} & 367 & 1875 & \textbf{138} & 264 & 506 & 884 & 1542 \\ \hline
No OLSD & 119 & \textbf{510} & 1457 & 2650 & \textbf{510} & 729 & 1060 & 1415 & 1975 \\ \hline
Only TTQ & 2880 & \textbf{5398} &  &  & \textbf{5398} &  &  &  &  \\ \hline
Only OLSD & 734 & \textbf{7791} & &  & \textbf{7791} &  &  &  &  \\ \hline
Only Masks & 8182 &  &  &  &  &  &  &  &  \\ \hline
No Intervention & 18590 &  &  &  &  &  &  &  &  \\ \hline
\end{tabular}\end{center}

\end{table}
\begin{table}[H]\caption{Peak quarantine size for various alternate values of $R_0$ and rates of infection from outside sources}
\label{outside2}
\begin{center}\begin{tabular}{|l||C{25.01pt}|C{25.01pt}|C{25.01pt}|C{25.01pt}||C{25.01pt}|C{25.01pt}|C{25.01pt}|C{25.01pt}|C{25.01pt}|}
\hline
 & \multicolumn{4}{c||}{$R_0$} & \multicolumn{5}{c|}{External Infections Per Day}\\
\hline
 & $2.8$ & \textbf{3.8} & $4.8$ & $5.8$ & \textbf{0.25} & $0.5$ & $1.0$ & $2.0$ & $4.0$ \\ \hline \hline
SI & 139 & \textbf{156} & 152 & 193 & \textbf{156} & 191 & 261 & 416 & 643 \\ \hline
No Mass Testing & 41 & \textbf{48} & 65 & 100 & \textbf{48} & 83 & 129 & 230 & 404 \\ \hline
No Distancing & 149 & \textbf{162} & 189 & 242 & \textbf{162} & 209 & 294 & 460 & 752 \\ \hline
No TTQ & $*$ & \textbf{$*$} & $*$ & $*$ & \textbf{$*$} & $*$ & $*$ & $*$ & $*$ \\ \hline
No Masks & 171 & \textbf{275} & 649 & 2472 & \textbf{275} & 403 & 617 & 881 & 1356 \\ \hline
No OLSD & 530 & \textbf{1760} & 3673 & 5082 & \textbf{1760} & 2124 & 2553 & 2798 & 3488 \\ \hline
Only TTQ & 6079 & \textbf{8673} &  &  & \textbf{8673} &  &  &  &  \\ \hline
Only OLSD & $*$ & \textbf{$*$} &  &  & \textbf{$*$} &  &  &  &  \\ \hline
Only Masks & $*$ &  &  &  &  &  &  &  &  \\ \hline
No Intervention & $*$ &  &  &  &  &  &  &  &  \\ \hline
\end{tabular}\end{center}

\end{table}

The final parameters we vary are the false positive rate and the false negative rate. Results are summarized in Tables \ref{fpr1} and \ref{fpr2}. The tables show a very strong dependence of the peak quarantine size on the FPR but much more muted relationships between FPR and total infections as well as FNR and both total infections and peak quarantine size. 
\begin{table}[H]\caption{Cumulative total infected for various alternate values of FPR and FNR}
\label{fpr1}
\begin{center}\begin{tabular}{|l||C{25.01pt}|C{27.01pt}|C{25.01pt}|C{25.01pt}|C{25.01pt}||C{25.01pt}|C{25.01pt}|C{25.01pt}|C{25.01pt}|C{25.01pt}|}
\hline
 & \multicolumn{5}{c||}{False Positive Rate} & \multicolumn{5}{c|}{False Negative Rate}\\
\hline
 & $10^{-4}$ & \textbf{0.001} & 0.005 & 0.01 & 0.02 & 0.015 & \textbf{0.03} & 0.06 & 0.12 & 0.24 \\ \hline \hline
SI & 45 & \textbf{44} & 41 & 43 & 40 & 44 & \textbf{44} & 42 & 43 & 45 \\ \hline
No Mass Testing & 54 & \textbf{46} & 50 & 46 & 50 & 49 & \textbf{46} & 49 & 52 & 45 \\ \hline
No Distancing & 49 & \textbf{48} & 49 & 43 & 44 & 49 & \textbf{48} & 52 & 51 & 49 \\ \hline
No TTQ & 87 & \textbf{83} & 81 & 74 & 89 & 82 & \textbf{83} & 78 & 89 & 78 \\ \hline
No Masks & 136 & \textbf{138} & 105 & 110 & 89 & 119 & \textbf{138} & 128 & 131 & 159 \\ \hline
No OLSD & 544 & \textbf{510} & 408 & 297 & 165 & 538 & \textbf{510} & 559 & 617 & 634 \\ \hline
\end{tabular}\end{center}

\end{table}
\begin{table}[H]\caption{Cumulative total infected for various alternate values of FPR and FNR}
\label{fpr2}
\begin{center}\begin{tabular}{|l||C{25.01pt}|C{27.01pt}|C{25.01pt}|C{25.01pt}|C{25.01pt}||C{25.01pt}|C{25.01pt}|C{25.01pt}|C{25.01pt}|C{25.01pt}|}
\hline
 & \multicolumn{5}{c||}{False Positive Rate} & \multicolumn{5}{c|}{False Negative Rate}\\
\hline
 & $10^{-4}$ & \textbf{0.001} & 0.005 & 0.01 & 0.02 & 0.015 & \textbf{0.03} & 0.06 & 0.12 & 0.24 \\ \hline \hline
SI & 77 & \textbf{156} & 444 & 791 & 1443 & 157 & \textbf{156} & 146 & 148 & 151 \\ \hline
No Mass Testing & 55 & \textbf{48} & 53 & 54 & 61 & 41 & \textbf{48} & 52 & 57 & 49 \\ \hline
No Distancing & 91 & \textbf{162} & 454 & 801 & 1462 & 166 & \textbf{162} & 164 & 159 & 163 \\ \hline
No TTQ & $*$ & \textbf{$*$} & $*$ & $*$ & $*$ & $*$ & \textbf{$*$} & $*$ & $*$ & $*$ \\ \hline
No Masks & 172 & \textbf{275} & 509 & 881 & 1487 & 267 & \textbf{275} & 268 & 278 & 295 \\ \hline
No OLSD & 1754 & \textbf{1760} & 1780 & 1982 & 2499 & 1746 & \textbf{1760} & 2009 & 1977 & 2008 \\ \hline
\end{tabular}\end{center}

\end{table}

\subsection{Analytical Model}

In this section, we present a simple theoretical model which gives further support for the observed results of our more general ABM. The theoretical model is essentially a discrete-time compartmental model regarding susceptible, infected, and removed individuals in each classroom as their own compartment. Because the compartments are not disjoint (i.e., because each student belongs to multiple classes), we include ``bookkeeping transmission'' between classes to properly reflect the effects of simultaneous membership in multiple classes. We find that the whole system can be conveniently understood in terms of an effective reproduction number $\tilde R$ given by
\begin{equation}
\tilde R :=  \gamma + c \frac{ \sum_\ell \frac{\alpha_\ell (S_\ell^0)^2}{1 - \alpha_\ell S_\ell^0}}{\sum_\ell \frac{S_\ell^0}{1 - \alpha_\ell S_\ell^0}}, \label{ourmodel}
\end{equation}
where $S_\ell^0$ is the initial susceptible population in class $\ell$, $\alpha_\ell$ is the per-person rate of transmission in class $\ell$ (so that an initial infected individual in class $\ell$ would be expected to infect $\alpha_\ell S_\ell^0$ total susceptible individuals), $c$ is the average number of classes per student\footnote{Formally, the parameter $c$ is a dimensionless parameter equal to the sum of all class sizes divided by the total number of students.}, and $\gamma$ is a dimensionless transmission rate for all non-classroom transmission. To avoid uncontrolled transmission within each classroom, it is necessary that $\alpha_\ell S_\ell^0 < 1$ for each $\ell$. Likewise, even if each individual classroom is below this threshold, uncontrolled transmission throughout the university is possible when $\tilde R > 1$.

Before we explain the derivation of \eqref{ourmodel}, we briefly explain why this formula further supports the idea that the largest classes have the biggest impact on disease dynamics.  In classroom $\ell$, the product $\alpha_\ell S_\ell$ corresponds to the ``local'' reproduction number within that classroom. The ratio
\[ \frac{ \sum_\ell \frac{\alpha_\ell (S_\ell^0)^2}{1 - \alpha_\ell S_\ell^0}}{\sum_\ell \frac{S_\ell^0}{1 - \alpha_\ell S_\ell^0}} \]
is a weighted average of these local reproduction number, with the weight in classroom $\ell$ being $S_\ell^0 / (1 - \alpha_\ell S_\ell^0)$. The weights favor larger classes because of the factor of $S_\ell^0$ in the numerator, and they also favor classes with higher local reproduction numbers because of the denominator $(1- \alpha_\ell S_\ell^0)$. Since in our case $\alpha_\ell S_\ell^0$ is an increasing function of class size, we expect \eqref{ourmodel} overall to be far more significantly impacted by larger classes than smaller ones.  Using the contact rates and transmission dynamics from our ABM, it is possible to compute $\tilde R$ explicitly under the assumption that only classes below a given threshold are meeting in person. Based on the parameters of our ABM, the parameters which most naturally correspond to it are $\gamma = 0.2$ and 
\[ \alpha_\ell = \begin{cases} 0.001392 & S_\ell^0 \leq 4, \\ 0.001392 + \frac{0.01728}{\sqrt{S_\ell^0}} & S_\ell^0 \geq 5. \end{cases}\]
These values roughly mirror the standard intervention in the presence of no residential contacts and with environmental contacts replaced by equal numbers of close academic and classroom contacts. If all classes of 30 or more are moved entirely online, the value of $\tilde R$ given by \eqref{ourmodel} is $0.52$ for a typical simulated distribution of class sizes; in our ABM we observe a median $R_0$ of $0.44$ in this situation. In particular, the theoretical model agrees with the ABM that this scenario is well below the threshold for uncontrolled exponential growth of infections.

A number of other values for different class size cutoffs are summarized in Table \ref{tmodel} below. The agreement is somewhat reduced as class size increases, but this is to be expected since our theoretical model is based on the assumption that the total number of cases is small, which becomes less appropriate as $\tilde R$ increases. 
\begin{table}[H]\caption{Computed $\tilde R$ versus simulation for various class size caps} \label{tmodel}
\begin{center}
\begin{tabular}{|c||c|c|}
\hline
Class Size Cap & Computed $\tilde R$ & Simulated $R_0$ \\ \hline \hline
10 & $0.29$ & $0.27$ \\ \hline
20 & $0.43$ & $0.33$ \\ \hline
30 & $0.52$ & $0.44$ \\ \hline
60 & $0.66$ & $0.54$ \\ \hline
90 & $0.92$ & $0.76$ \\ \hline
\end{tabular} \end{center}
\end{table}

The model equations are as follows. We will treat time as a discrete variable and let $S_k^{n}$ represent the number of susceptible students in class $k$ on day $n$; the difference $I_k^n := S_k^{n-1} - S_{k}^n$ represents the number of new cases in class $k$ on day $n$.  We assume that the number of new cases in class $k$ which are due to individuals infected in class $\ell$ is roughly equal to $S_k^{n} \sum_{d=1}^{14} R_{k\ell}^d I_{\ell}^{n-d+1}$ in the limit of very small transmission proportions $R_{k\ell}^d$. We also assume that there is an external source causing  $F_{k}^{n+1} S_k^n$ additional infections in the limit of small $F_k^{n+1}$.
Our equations are
\begin{equation} S^{n+1}_k := S_{k}^n \exp \left( - F_k^{n+1} - \sum_{d=1}^{14} \sum_{\ell} R_{k\ell}^{d} I_\ell^{n-d+1} \right) .  \label{themodel} \end{equation}
Note that when $F_{k}^{n}$ is identically zero, \eqref{themodel} admits solutions which are constant for all time. To cleanly identify physically-realistic initial conditions, we define $S^{n}_k := S^{0}_k$ for all $n < 0$, which effectively means that we are identifying those solutions which existed in constant equilibrium before the introduction of external infections on days $n \geq 0$.
We note that in the regime of small outbreaks, each $I_\ell^{n-d+1}$ will be order $1$, so in the limit as $F_{k}^{n+1}, R_{k\ell}^{d} \rightarrow 0$, we have a valid expansion of the exponential \eqref{themodel}
\begin{equation}
 S_k^{n+1}  = S_k^n \left( 1 - F_k^{n+1} - \sum_{d=1}^{14}\sum_{\ell} R_{k \ell}^d I_\ell^{n-d+1} \right) + \text{(higher order terms)}. \label{expansion}
 \end{equation}
Even though we are in a perturbative regime, it is helpful to work directly with \eqref{themodel} rather than the expansion \eqref{expansion} because \eqref{themodel} admits useful conserved quantities which allow for an easier understanding of the limiting behavior as $n \rightarrow \infty$. Specifically, fixing
\[ \Phi_k^{n} := S_k^n \exp \left( - \sum_{d=1}^{14} \sum_{\ell} R_{k \ell}^d S_\ell^{n-d} \right) \]
and using the identity $I_\ell^{n-d+1} = S_\ell^{n-d} - S_\ell^{n-d+1}$ gives the identity
\[ \Phi_k^{n+1} = \Phi_k^{n} e^{- F_k^{n+1}}.\]
When the $F_k^n$ are nonnegative, it is easy to see by induction that the quantities $S_k^n$ and $\Phi_k^n$ are strictly positive and nonincreasing in $n$; thus the limits $S_k := \lim_{n \rightarrow \infty} S_k^n$ and $\Phi_k := \lim_{n \rightarrow \infty} \Phi_k^n$ exist and
\[ S_k \exp \left( - \sum_{d=1}^{14} \sum_\ell R^{d}_{k \ell} S_\ell \right) = S^0_k \exp \left( - \sum_{n=1}^\infty F_k^n - \sum_{d=1}^{14} \sum_\ell R^{d}_{k \ell} S_\ell^0 \right).  \]
Among other features, we see that the temporal structure of the transmission parameters $R^d_{k\ell}$ will affect the temporal dynamics but only their sum is needed to understand the limiting behavior; for convenience, we simply let $R_{k\ell} := \sum_{d=1}^{14} R^d_{k \ell}$ and $F_k := \sum_{n=1}^\infty F_k^n$ and work directly with these quantities. This means that the limiting values $S_k$ satisfy 
\begin{equation}
 \Phi_k = S_k \exp \left( - \sum_\ell R_{k \ell} S_\ell \right) = S_k^0 \exp \left( - F_k - \sum_{\ell} R_{k \ell} S_\ell^0 \right) = e^{-F_k} \Phi_k^0. \label{solution}
 \end{equation}
provided that $F_k < \infty$ for each $k$, the right-hand side of \eqref{solution} is always strictly positive, which means that the limit values $S_k$ are also strictly positive.

We will model the transmission proportions $R_{k \ell}$ as a sum of two terms: $R_{k \ell} = \alpha_k \delta_{k\ell} + \beta T^{-1}$, where $\delta_{k\ell}$ is the Kronecker delta. Here each $\alpha_k$ models transmission within classroom $k$, and $\beta T^{-1}$ is a parameter which will account for broad transmission as well as necessary ``bookkeeping'' reflect the fact that individual students are enrolled in multiple courses (here $T$ is simply the total number of students times the average number of courses per student).  The equation \eqref{solution} becomes
\begin{equation} S_k e^{- \alpha_k S_k} e^{- \frac{\beta}{T} \sum_{\ell} S_\ell} = e^{- F_k} S_k^0 e^{- \alpha_k S_k^0} e^{- \frac{\beta}{T} \sum_{\ell} S_\ell^0}. \label{ourrates} \end{equation}
This identity can be rewritten as
\[ S_k e^{-\alpha_k S_k} = e^{- F_k} S_k^0 e^{- \alpha_k S_k^0} e^{- \frac{\beta}{T} \sum_{\ell} (S_\ell^0-S_\ell) }; \]
and treating the right-hand side as an arbitrary fixed, small perturbation of $S_k^0 e^{-\alpha_k S_k^0}$, the equation will admit a solution $S_k$ near to $S_k^0$ varying smoothly with the perturbation if and only if $\alpha_k S_k^0 < 1$, simply because the function $x e^{-\alpha x}$ has a critical point at $x = 1/\alpha$. Thus, to prevent a localized, uncontrolled outbreak in class $k$, it is necessary that $\alpha_k S_k^0 < 1$.  Returning to \eqref{ourrates}, we must also compute the Jacobian matrix $\partial \Phi / \partial S$. To that end, we compute:
\[ \frac{\partial}{\partial S_\ell} S_k e^{- \alpha_k S_k} e^{- \frac{\beta}{T} \sum_{\ell} S_\ell} = \left( \left( \frac{1}{S_k} - \alpha_k \right) \delta_{k\ell}- \frac{\beta}{T} \right) S_k e^{- \alpha_k S_k} e^{- \frac{\beta}{T} \sum_{\ell} S_\ell}. \]
Treating the right-hand side of \eqref{ourrates} as a small perturbation of $\Phi_k^0$ for each $k$, we also know that the solutions $S_k$ will vary smoothly in this perturbation only when $\beta$ is sufficiently small that the Jacobian matrix is invertible. For convenience, we give the matrix entries just computed names: if $M_{k\ell} := \partial \Phi_k / \partial S_\ell$, then
\[ M_{k\ell} := \Phi_k^0 \left[ \left( S_k^{-1} - \alpha_k \right) \delta_{k \ell} - \beta T^{-1} \right]. \]
One can see by an explicit calculation that $M$ is invertible as long as
\begin{equation}
\frac{\beta}{T} \sum_{\ell} \frac{S_\ell}{1- \alpha_\ell S_\ell} < 1 \label{stable0}
\end{equation}
and that $M$ has inverse
\[ M^{-1}_{\ell j} = \left[ \frac{S_\ell \delta_{\ell j}}{1-\alpha_\ell S_\ell} + \frac{\beta}{T - \beta \sum_{m} \frac{S_m}{1- \alpha_m S_m}} \frac{S_\ell S_j}{(1 - \alpha_\ell S_\ell) (1 - \alpha_j S_j)} \right] \frac{1}{\Phi_j^0}. \]
(We note that the solution map must be discontinuous when the left-hand side of \eqref{stable0} is strictly positive since in this case $M^{-1}$ has negative entries, which, if the map were continuous, would mean that increasing $F_k$ for certain $k$ would decrease the number of infected in some classroom $k'$, which can be easily ruled out by an elementary analysis of the equations \eqref{themodel}.) 
In particular, in the limit of small $F_k$'s, one has the linear approximation
\[ S_k \sim S_k^0 - \frac{S_k^0 F_k}{1 - \alpha_k S_k^0} - \frac{\beta}{T - \beta \sum_{m} \frac{S_m^0}{1- \alpha_m S_m^0}} \frac{S_k^0}{1 - \alpha_k S_k^0} \sum_j \frac{ F_j S_j^0}{1 - \alpha_j S_j^0}. \]
To correctly choose $\beta$, we momentarily fix each $\alpha_k = 0$ and we let $F_{k_0} = \epsilon (S_{k_0}^0)^{-1}$ for a single index $k = k_0$ and set $F_{k} = 0$ otherwise. This represents the introduction of an external infection into class $k$ of size $\epsilon$. We get
\begin{equation} S_k \sim S_k^0 - \epsilon \delta_{kk_0} - \frac{\beta}{(1-\beta)T} S_k^0 \epsilon, \label{crosscontam} \end{equation}
and summing over $k$ gives
\[ T - \sum_{k} S_k  \sim \epsilon \left[ 1 + \frac{\beta}{1-\beta} \right] = \frac{\epsilon }{1-\beta}. \]
If the model were capturing only broad contact across the entire university, we would expect the right-hand side to equal $c \epsilon (1-\gamma)^{-1}$, where $c$ counts the average number of courses a student takes. Thus we should fix $1 - \beta = c^{-1} (1-\gamma)$. We also see from \eqref{crosscontam} that the term
\[ \frac{\beta}{(1-\beta) T} S_k^0 \epsilon \]
does not depend on the initial classroom $k_0$ and is proportional to $S_k^0$, which means that we are effectively assuming that students choose their courses independently with probability proportional to class size.

 Making the choice of $\beta$ identified above and rewriting \eqref{stable0} at the initial data $S_k^0$ gives the stability criterion
\[ \frac{1 - c^{-1}(1-\gamma)}{T} \sum_{\ell} \frac{S_\ell^0}{1 - \alpha_\ell S_\ell^0} < 1, \]
and using the fact that $T = \sum_{\ell} S_\ell^0$ to rewrite this condition gives the equivalent condition
\[ \tilde R = \gamma + c \frac{\sum_{\ell} \frac{\alpha_\ell (S_\ell^0)^2}{1 - \alpha_\ell S_\ell^0}}{\sum_{\ell} \frac{S_\ell^0}{1 - \alpha_\ell S_\ell^0}} < 1. \]
As above, if we let $F_k = \epsilon $ for each $k$ and let $\epsilon \rightarrow 0$, we have
\[ \sum_k (S_k^0 - S_k) \sim \frac{T c \epsilon}{1 - \tilde R}. \]
This formula is what motivates our description of $\tilde R$ as an effective reproduction number. Setting $F_k = \epsilon$ for each $k$ corresponds to a total of $T \epsilon$ infections from outside sources, the factor of $c$ merely reflects the bookkeeping requirement that one infection from an external source contributes an amount of $c$ when counted with multiplicity across all classes. Therefore one sees that $\tilde R$ is an effective reproduction number for the system when individuals are subjected to a small external source of infections.

\end{document}